\documentclass[final,3p,times]{elsarticle}

\usepackage{booktabs}   %% For formal tables:
\usepackage{subcaption} %% For complex figures with subfigures/subcaptions
\usepackage[utf8]{inputenc}

\usepackage{xcolor}
\usepackage{natbib}
\usepackage{graphicx}
\usepackage{enumitem}

\usepackage[ruled,vlined,linesnumbered]{algorithm2e} %vlined,
\usepackage{tikz}
\usepackage{tikzscale}
\definecolor{darkgreen}{RGB}{0,104,0}

\usepackage{amsthm}
\usepackage{amsmath}
\usepackage{amssymb}
\usepackage{amsfonts}

\usepackage{hyperref}
\usepackage{multirow}

\usepackage{caption}
\usepackage{subcaption}

\newtheorem{theorem}{Theorem}[section]

\theoremstyle{definition}
\newtheorem{definition}{Definition}[section]
\theoremstyle{definition}
\newtheorem{example}{Example}[section]
\usepackage{xcolor}

\newcommand{\od}{\mathbb{O}}  %%% for the k-order, distinguish with O(n).

\newcommand{\enqueue}{{\it enqueue}}

\newcommand{\pre}{{\it{pre}}}

\newcommand{\group}{{\it group}}
\newcommand{\lock}{{\it lock}}
\newcommand{\Next}{{\it{next}}}
\newcommand{\Push}{\it{push}}
\newcommand{\Pop}{\it{pop}}
\newcommand{\TOP}{\it{top}}

\newcommand{\version}{\alpha}
\newcommand{\Version}{A}
\newcommand{\live}{{\it{live}}}

\newcommand{\alg}[1]{\textsc{#1}} %%%% all algorithm names use. 
\newcommand{\omorder}{\texttt{Order}~}
\newcommand{\ominsert}{\texttt{Insert}~}
\newcommand{\omdelete}{\texttt{Delete}~}

\usepackage{xcolor}
\newcommand{\mytodo}[1]{\textcolor{red}{#1}}

\journal{Journal of Parallel and Distributed Computing}
\begin{document}
\begin{frontmatter}
\title{New Concurrent Order Maintenance Data Structure}
% \author{Bin Guo}
% \author{Emil Sekerinski}
% \date{\today}

\author[label1]{Bin Guo\corref{cor1}}
\ead{binguo@trentu.ca}
\cortext[cor1]{Corresponding author}
\affiliation[label1]{organization={Department of  Computing \& Information Systems, Trent University},%Department and Organization
            addressline={1600 West Bank Drive}, 
            city={Peterborough},
            postcode={K9L 0G2}, 
            state={ON},
            country={Canada}}

 \author[label2]{Emil Sekerinski}
 \ead{emil@mcmaster.ca}
\affiliation[label2]{organization={Department of Computing and Software, McMaster University},%Department and Organization
            addressline={1280 Main Street West}, 
            city={Hamilton},
            postcode={L8S 4L8}, 
            state={ON},
            country={Canada}}  

\begin{abstract}
The \emph{Order-Maintenance} (OM) data structure maintains a total order list of items for insertions, deletions, and comparisons. As a basic data structure, OM has many applications, such as maintaining the topological order, core numbers, and truss in graphs, and maintaining ordered sets in Unified Modeling Language (UML) Specification. The prevalence of multicore machines suggests parallelizing such a basic data structure. This paper proposes a new parallel OM data structure that supports insertions, deletions, and comparisons in parallel. Specifically, parallel insertions and deletions are synchronized by using locks efficiently, which achieve up to $7$x and $5.6$x speedups with $64$ workers. One big advantage is that the comparisons are lock-free so that they can execute highly in parallel with other insertions and deletions, which achieve up to $34.4$x speedups with $64$ workers. Typical real applications maintain order lists that always have a much larger portion of comparisons than insertions and deletions. For example, in core maintenance, the number of comparisons is up to 297 times larger compared with insertions and deletions in certain graphs. This is why the lock-free order comparison is a breakthrough in practice. 
\end{abstract}

% %% Keywords
% %% comma separated list
% \keywords{order maintenance, parallel, multi-core, shared memory, compare-and-swap, lock-free, amortized constant time, core maintenance}  %% \keywords are mandatory in final camera-ready submission

\iffalse 
\begin{highlights}
\item The Order Maintenance data structure has amortized $O(1)$ time for Insert, Delete, and Order operations.  
\item Execute efficiently on multicore shared memory machines in parallel.
\item It has applications such as maintaining the topological order of vertices in parallel.
\end{highlights}
\fi 


\end{frontmatter}

\section{Introduction}
The well-known \emph{Order-Maintenance} (OM) data structure \cite{dietz1987two,bender2002two,utterback2016provably} maintains a total order of unique items in an order list, denoted as $\od$, by following three operations:
\begin{itemize}%[leftmargin=*] %[noitemsep,topsep=0pt,leftmargin=* ] % , leftmargin=* 
    \item $\texttt{Order}(x, y)$: determine if $x$ precedes $y$ in the ordered list $\od$, denoted as $x \preceq y$, supposing both $x$ and $y$ are in $\od$.
    \item $\texttt{Insert}(x, y)$: insert a new item $y$ after $x$ in the ordered list $\od$, supposing $x$ is in $\od$ and $y$ is not in $\od$.
    \item $\texttt{Delete}(x)$: delete $x$ from the ordered list $\od$, supposing $x$ is in $\od$.
\end{itemize}

\paragraph{Application}
As a fundamental cohesive subgraph model, the \emph{$k$-core}~\cite{bz2003,cheng2011efficient,khaouid2015k,montresor2012distributed,wen2016efficient} is defined as the maximal subgraph such that all vertices have degrees at least $k$. The \emph{core number} of a vertex is defined as its maximum value of $k$. After core numbers of vertices are computed in linear time by peeling steps~\cite{bz2003}, it is time-consuming to recalculate the core numbers when new edges are inserted or old edges are removed for dynamic graphs. In this case, \emph{core maintenance} algorithms~\cite{sariyuce2013streaming,wu2015core,Zhang2017,guo2022simplified} are proposed to efficiently update the core numbers of vertices, which avoids traversing the whole graph. In~\cite{Zhang2017,guo2022simplified}, the \emph{$k$-order} of all vertices are used for core maintenance, where $u$ precedes $v$ in the peeling steps of core decomposition for all vertices $u,v$ in the graph. The idea is that after inserting an edge, the related vertices $w$ are traversed in $k$-order; we can safely skip $w$ if $w$ has candidate in-degree plus remaining out-degree less than $k+1$, and thus the number of traversed vertices can be significantly reduced. 

\begin{figure}[!htb]
\centering
\includegraphics[width=0.7\linewidth]{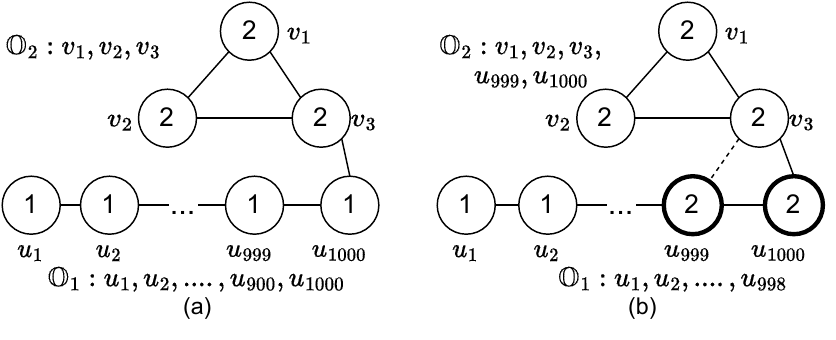} %[width=\linewidth]
\caption{An example of core maintenance by using OM data structure. The letters outside circles are vertices' IDs, the numbers inside circles are vertices' core numbers, and the $O_k$ is the $k$-order of vertices with the core number $k$ maintained by OM data structure.  }
\label{fig:cm}
\end{figure}

\begin{example}
    %In Figure~\ref{fig:cm}, we show an example graph that maintains the core numbers. 
    Figure~\ref{fig:cm}(a) shows all the core numbers of vertices computed by peeling. For example, at first, $u_1$ obtains a core number as $1$ since its degree is $1$; after peeling $u_1$, $u_2$ gets core numbers as $1$ since its degree is reduced to $1$; continually, $u_3$ to $u_{1000}$ gets the core numbers as $1$. Here, the $k$-order $O_1$ is the order of vertices that obtain their core numbers.
    %Similarly, $v_1$ to $v_3$ obtain core numbers as $2$.      
    In Figure~\ref{fig:cm}(b), inserting a new edge $(u_{999}, v_3)$ may cause the core number of some vertices to increase from $1$ to $2$. We traverse the affected vertices starting from $u_{999}$ by $O_1$. In this case, we traverse only two vertices, $u_{999}$ and $u_{1000}$, and find that their core numbers can increase from $1$ to $2$. We can see a great number of vertices, $u_1$ to $u_{998}$, are avoided to the traversed, which achieves high performance for maintaining the core numbers of vertices. Finally, to maintain the $k$-order, $u_{999}$ and $u_{1000}$ are removed from $\od_1$ and appended to $\od_2$. 
\end{example}

% In~\cite{marchetti1996maintaining, haeupler2012incremental}, given a directed graph, a \emph{topological order} of all vertices are maintained, where $u$ precedes $v$ in topological order for all edges $(u, v)$ in the graph. 
% In~\cite{Zhang2017,guo2022simplified}, given a undirected graph, a \emph{$k$-order} of all vertices are used for core maintenance, where $u$ is precedes $v$ in the peeling steps of core decomposition~\cite{bz2003,cheng2011efficient,khaouid2015k,montresor2012distributed,wen2016efficient} for all edges $(u, v)$ in the graph. 

Due to the prevalence of the multicore shared-memory architecture, many parallel core maintenance approaches were proposed~\cite{hua2019faster,Jin2018,wang2017parallel,guo2022parallel}. As one of the important uses in~\cite{guo2022parallel}, our concurrent OM data structure is used to maintain the $k$-order in parallel, when multiple edges are inserted simultaneously. 
Such an idea about core maintenance can be applied to truss maintenance~\cite{zhang2019unboundedness}. Similarly, the OM data structure can be used to maintain the \emph{topological order} of vertices in directed acyclic graphs after inserting or removing edges, where for every directed edge $(u, v)$ we have $u$ come before $v$ in order~\cite{bender2009new,marchetti1996maintaining}.     

Additionally, ordered sets are widely used in Unified Modeling Language (UML) Specifications~\cite{martin2003agile}, e.g., a display screen (an OS's representation) has a set of windows, but furthermore, the set is ordered, so do the ordered bag and sequence.
Nowadays, shared-memory multi-core machines are widely used, which motivates the efficient parallelization of the above algorithms. 
%where different cores have access to a shared global memory for synchronization. Based on such architecture, many parallel algorithms are proposed. 
As a building block, it immediately suggests itself parallelizing the OM data structure.  
%including the parallelism of topological order maintenance and core maintenance. Thus, it is an urgent issue to parallelize the OM data structure on shared-memory multi-core machines. 
%%%% add a application for order-based core maintenance.

%%%%%%%%%%%%%%%%%%%%%%%%%%%%%%%%%%%%%
% add an example to here. use k-order to maintain the core maintenance. To show that our k-order is important. 
% and to show that most of order operation are order rather than insert or delete. 
%%%%%%%%%%%%%%%%%%%%%%%%%%%%%%%%%%%%%

\paragraph{Sequential} 
In the sequential case, the OM data structure has been well studied. 
The naive idea is to use a balanced binary search tree \cite{cormen2022introduction}. All three operations can be performed in $O(\log N)$ time, where there are at most $N$ items in the ordered list $\od$. 
In~\cite{dietz1987two,bender2002two}, the authors propose an OM data structure that supports all three operations in $O(1)$ time. The idea is that all items in $\od$ are linked as a double-linked list. 
Each item is assigned a label to indicate its order. 
We can perform the \texttt{Order} operation by comparing the labels of two items by $O(1)$ time. Also, the \texttt{Delete} operation also costs $O(1)$ time without changing other labels.
For the \texttt{Insert}$(x, y)$ operation, $y$ can be directly inserted after $x$ with $O(1)$ time, if there exists label space between $x$ and $x$'s successors; otherwise, a \emph{relabel} procedure is triggered to rebalance the labels, which costs amortized $O(\log N)$ time per insertion.
After introducing a list of sublists structure, the amortized running time of the relabel procedure can be optimized to $O(1)$ per insertion. Thus, the \texttt{Insert} operation has $O(1)$ amortized time. 

\paragraph{Parallel}
In this paper, we present a new concurrent OM data structure. 
In terms of parallel \texttt{Insert} and \texttt{Delete} operations, we use locks for synchronization without allowing interleaving. 
In the average case, there is a high probability that multiple \texttt{Insert} or \texttt{Delete} operations occur in different positions of $\od$ so that these operations can execute completely in parallel.
% In the worst case, if the \texttt{Insert} or \texttt{Delete} operation occurs in the same position after one item $x$, it is reduce to sequential because of sequentially locked items.
% Typically, such worst-case has low probability to happen in real applications, e.g, many items are inserted at the head or tail of the ordered list $\od$. 
For the \texttt{Order} operation, we adopt a lock-free mechanism, which allows executing completely in parallel for any pair of items in $\od$.
To implement the lock-free \texttt{Order}, we devise a new algorithm for the \texttt{Insert} operation that always maintains the \emph{Order Snapshot} for all items, even if many relabel procedures are triggered. 
Here, the Order Snapshot means the labels of items indicate their order correctly. 
As \texttt{Insert} operations always maintain the Order Snapshot, we do not need to lock a pair of items when comparing their labels in parallel. 
In other words, lock-free \texttt{Order} operations are based on \texttt{Insert} operations that preserve the Order Snapshot. 

\begin{figure}[!htb]
\centering
\includegraphics[width=0.7\linewidth]{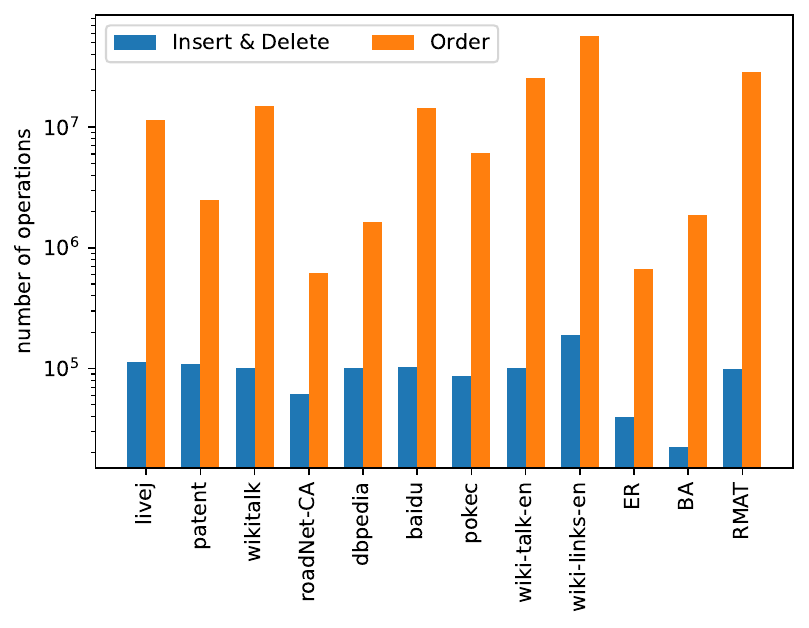} %[width=\linewidth]
\caption{The number of OM operations for core maintenance by inserting $100,000$ random edges into each graph.}
\label{fig:om-core-maint}
\end{figure}

Our new parallel lock-free \texttt{Order} operation is a breakthrough for real applications. Typically, for the OM data structure, a large portion of operations is comparing the order of two items. 
For example, Figure~\ref{fig:om-core-maint} shows the number of OM operations (y-axis) for the core maintenance in~\cite{guo2022simplified} by randomly inserting $100,000$ edges over $12$ tested data graphs (x-axis).
%where the core numbers are already constructed for these tested graphs. 
We observe that the number of \texttt{Order} operations is magnitudes larger than the number of \texttt{Insert} and \texttt{Delete} operations, e.g., for the RMAT graph by a factor of 287.
The reason is that graphs tend to have more edges than vertices. The crucial advantage of our parallel \texttt{Order} operation is that it can execute completely in parallel without locking items, which is essential when trying to parallelize algorithms like core maintenance.

We analyze our parallel OM operations in the work-depth model~\cite{cormen2022introduction,shun2017shared}, where the work, denoted as $\mathcal W$, is the total number of operations that are used by the algorithm, and the depth, denoted as $\mathcal D$, is the longest length of sequential operations~\cite{jeje1992introduction}. The expected running time is $O(\mathcal W/ \mathcal P + \mathcal D)$ by using $\mathcal P$ workers with load balancing among those. In particular, the work and depth terms are equivalent for sequential algorithms. 

\begin{table}[!htb]
\centering
%\small
\begin{tabular}{l|c c c|c c c}
\hline
   Parallel    & \multicolumn{3}{ c}{Worst-case ($O$) } &  \multicolumn{3}{|c}{Best-case ($O$)}   \\ % \cline{2-5}
  operation     & $\mathcal W$ & $\mathcal D$ & time & $\mathcal W$ & $\mathcal D$ & time\\ \hline %& space\\  \hline
 \texttt{Insert}&     $m^\dagger $ & $m^\dagger $ & $\frac{m}{\mathcal P} + m^\dagger $  & $m^\dagger$   & $1^\dagger$ & $\frac{m}{\mathcal P}^\dagger$ \\%& $O(N)$ \\
 \texttt{Delete} & $m$          &$m$     &   $\frac{m}{\mathcal P} + m$  & $m$    & $1$       &   $\frac{m}{\mathcal P}$       \\%& $O(N)$ \\
 \texttt{Order}&  $m$         & $1$       &  $\frac{m}{\mathcal P}$      & $m$    & $1$        &   $\frac{m}{\mathcal P}$      \\%& $O(N)$ \\
 \hline
\end{tabular}
\caption{The worst-case and best-case work, depth complexities of parallel OM operations, where $m$ is the number of operations executed in parallel, $\mathcal P$ is the total number of workers, and $^\dagger$ is the amortized complexity. }
\label{tab:comp}
\end{table}  

Table \ref{tab:comp} compares the worst-case and best-case work and depth complexities for the three OM operations when running the $m$ operations of the same kind in parallel. 
In the best case, all three operations have $O(m)$ work and $O(1)$ depth. 
However, \texttt{Insert} has worst-case $O(m)$ work and $O(m)$ depth; such a worst-case is easy to construct by inserting $m$ items into the same position of $\od$, and thus all insertions are reduced to running sequentially.
The \texttt{Delete} operation also has worst-case $O(m)$ work and $O(1)$ depth; but such a worst case only happens when all deletions cause a blocking chain, which has a very low probability. 
Especially, since the \texttt{Order} is lock-free, it always has $O(m)$ work and $O(1)$ depth in the worst and best cases. 
This is why \texttt{Order} operations run in parallel always have a great speedup for multicore machines. 
The lock-free \texttt{Order} operation is an important contribution of this work.

We conduct extensive experiments on a 64-core machine over a variety of test cases to evaluate the parallelism of the new parallel OM data structure. With 64 workers our parallel \texttt{Insert} and \texttt{Delete} achieve up to 7x and 5.6x speedups; our parallel \texttt{Order} achieves up to 34.4x speedups.

\iffalse
The main contributions of this work are summarized below:
\begin{itemize} % [noitemsep,topsep=0pt]
\item We propose a parallelized OM data structure that supports the \texttt{Insert} and \texttt{Delete} operations in parallel, which is synchronized by using lock. In average-case, both operations can execute highly in parallel when they occurs in different positionof the ordered list $\od$. 

\item We propose a new \texttt{Order} operation that can run highly in parallel, as we use STM instead of lock for synchronization.  

\item 

\end{itemize}
\fi 

%%%%???
The rest of this paper is organized as follows. The related work is in Section 2. The preliminaries are given in Section 3. Our parallel OM data structure is discussed in Section 4.
We conduct experimental studies in Section 5 and conclude this work in Section 6.

\section{Related Work}
In~\cite{dietz1982maintaining}, Dietz proposes the first order data structure, with \texttt{Insert} and \texttt{Delete} having $O(\log n)$ amortized time and \texttt{Order} having $O(1)$ time. 
In~\cite{tsakalidis1984maintaining}, Tsakalidis uses BB[$\alpha$] trees to improve the update bound to $O(\log n)$ and then to $O(1)$ amortized time.
In~\cite{dietz1987two}, Dietz et al. propose the fastest order data structure, which has \texttt{Insert} in $O(1)$ amortized time, \texttt{Delete} in $O(1)$ time, and \texttt{Order} in $O(1)$ time. 
In~\cite{bender2002two}, Bender et al. propose significantly simplified algorithms that match the bounds in~\cite{dietz1987two}.

A special case of OM is the \emph{file maintenance} problem~\cite{dietz1987two,bender2002two}, which is to store $n$ items in an array of size $O(n)$. File maintenance has four operations, i.e., insert, delete, scan-right (scan next $k$ items starting from $e$), and scan-left (analogous to scan-right).

 %%%% add history of this problem from wikipedia. 
For the parallel or concurrent OM data structure, there exists little work~\cite{gilbert2003concurrent,utterback2016provably} to the best of our knowledge. 
In~\cite{gilbert2003concurrent}, the order list is split into multiple parts and organized as a B-tree, which sacrifices the $O(1)$ time for three operations; also, the relevant nodes in the B-tree are locked for synchronization.  
In~\cite{utterback2016provably}, a parallel OM data structure is proposed specifically for \emph{series-parallel (SP)} maintenance, which identifies whether two accesses are logically independent. 
Several parallelism strategies are present for the OM data structure combined with SP maintenance. We apply the strategy of splitting a full group into our new parallel OM data structure.

\iffalse
The \alg{Order} algorithm\cite{Zhang2017} is efficient for core maintenance in graphs in the sequential case. The main idea is that a \emph{$k$-order} is explicitly maintained among vertices such that $u\preceq v$ for each pair of vertices in a graph $G$. Here, a $k$-order, $(v_1 \preceq v_2 \preceq .... \preceq v_n)$, for each vertex~$v_i$ in a graph $G$, is an order that the core number determined by a core decomposition algorithm~\cite{bz2003,cheng2011efficient,khaouid2015k,montresor2012distributed,wen2016efficient}, e.g., \alg{BZ} algorithm \cite{bz2003}. 
The $k$-order is implemented by double linked lists combined with min-heap, which requires $O(\log n)$ time for each \emph{order operation} (like inserting an item between two vertices, removing one vertex, and comparing the order of two vertices). 
In~\cite{guo2022simplified}, the OM data structure is adopted to maintain the $k$-order, which only requires $O(1)$ amortized time for each order operation. To parallelize the \alg{Order} algorithm, we must first parallelize the OM data structure for maintaining the $k$-order in parallel.
\fi 

\section{Preliminaries}
In this section, for the OM data structure, we revisit the detailed steps of the sequential version \cite{bender2002two,dietz1987two,utterback2016provably}. This is the background to discuss the parallel version in the next section. 

The idea is that items in the total order are assigned labels to indicate the order.
Typically, each label can be stored as an $O(\log N)$ bits integer, where $N$ is the maximal number of items in $\od$.
Assume it takes $O(1)$ time to compare two integers. 
The \omorder operation requires $O(1)$ time by comparing labels; also, the \omdelete operation requires $O(1)$ time since after deleting one item all other labels of items are not affected.

In terms of the \ominsert operation, efficient implementations provide $O(1)$ amortized time. 
First, a \emph{two-level} data structure~\cite{utterback2016provably} is used. That is, each item is stored in the bottom-list, which contains a \emph{group} of consecutive elements; each group is stored in top-list, which can contain $\Omega(\log N)$ items. 
Both the top-list and the bottom-list are organized as double-linked lists, and we use $x.\pre$ and $x.\Next$ to denote the predecessor and successor of $x$, respectively.    
Second, each item $x$ has a top-label $L^t(x)$, which equals to $x$'s group label denoted as $L^t(x) = L(x.\group)$, and bottom-label $L_b(x)$, which is $x$'s label.  
Integer $L^t$ is in the range $[0, N^2]$ and integer $L_b$ in the range $[0, N]$.

%%%%% the initial case should not be sepcial. 
Initially, there can be $N'$ items in $\od$ ($N' \leq N$), which are contained in $N'$ groups, separately. Each group is assigned a top-label $L$ with an $N$ gap between neighbors, and each item is assigned a bottom-label $L_b$ as $\lfloor N/2\rfloor$.

\begin{definition}[Order Snapshot]
  The OM data structure maintains the {Order Snapshot} for $x$ precedes $y$ in the total order, denoted as 
  $\forall x, y \in \od: x \preceq y \equiv L^t(x) < L^t(y)~\lor~(L^t(x) = L^t(y) \land L_b(x) < L_b(y))$
  \label{def:order}
 \end{definition}
The OM data structure maintains the Order Snapshot defined in Definition~\ref{def:order}. 
In order words, to determine the order of $x$ and $y$, we first compare their top-labels (group labels) of $x$ and $y$; if they are the same, we continually compare their bottom labels.

\subsection{Insert} 
The operation $\texttt{Insert}(\od, x, y)$ is implemented by inserting $y$ after $x$ in $x$'s bottom-list, assigning $y$ the label $L_b(y) = \lfloor(L_b(x.next) - L_b(x))/2\rfloor$, and setting $y$ in the same group as $x$ with $y.group = x.group$ such that $L^t(y) = L^t(x)$.
If $L_b(x.\Next) -L_b(x) > 1$, $y$ can successfully obtain a new label, then the insertion is complete in $O(1)$ time. 
Otherwise, the $x$'s group is \emph{full}, which triggers a \emph{{relabel}} operation. 
The relabel operation has two steps. First, the full group is split into many new groups, each of which contains at most $\frac{\log N}{2}$ items, and new labels $L_b$ are uniformly assigned for items in new groups.
Second, newly created groups are inserted into the top-list, if new group labels $L^t$ can be assigned. Otherwise, we have to \emph{{rebalance}} the group labels. That is, from the current group $g$, we continuously traverse the successors $g'$ until $L(g')-L(g)> j^2$, where $j$ is the number of traversed groups. 
Then, new group labels can be assigned to groups between $g$ and $g'$ with a $\frac{L(g')-L(g)}{j}$ gap, in which newly created groups can be inserted.

There are three important features in the implementation \texttt{Insert}: (1) each group, stored in the top-list, contains $\Omega(\log N)$ items, so that the total number of insertions is $O(N/\log N)$; (2) the amortized cost of splitting groups is $O(1)$ per insert; (3) the amortized cost of inserting a new group into the top-list is $O(\log N)$ per insertion. Thus, each \ominsert operation only costs amortized $O(1)$ time.
%%%% the locked should v1, v5, v6 ???
%%%%% also add subtitle. 
\begin{figure}[!htb]
\centering
\includegraphics[width=0.7\linewidth]{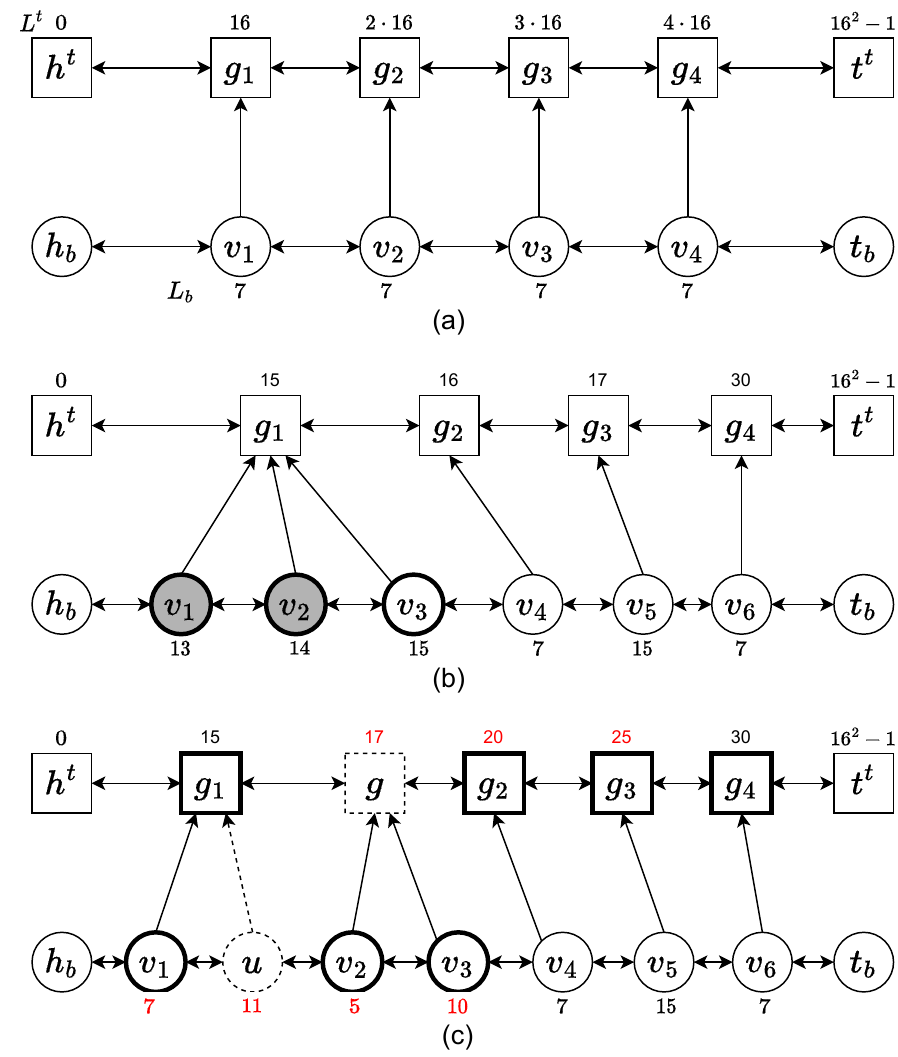} %[scale=0.5] width=\linewidth
%\caption{A example of the OM data structure with $N=16$. (a) the initial state; (b) the intermediate state after multiple insert and delete operations, where there has no label space after $v_1$ and thus $g_1$ is full; (c) after relabeling, insert $u$ between $v_1$ and $v_2$. }
\caption{An example of the OM data structure with $N=16$. 
The squares are groups located in a single double-linked top-list with a head $h^t$ and a tail $t^t$. The circles are items with pointers to their own groups, located in a single double-linked bottom-list. The beside numbers are corresponding labels to items and groups. 
(a) the initial state of $\od = \{v_1, v_2, v_3, v_4\}$. 
(b) the intermediate state of $\od = \{v_1, v_2, v_3, v_4, v_5, v_6\}$.
(c) after inserting a new item $u$ (dashed cycles) after $v_1$, we get $\od=\{v_1, u, v_2, v_3, v_4, v_5, v_6\}$ by inserting a new group $g$ (dashed square).
}
% ; the shaded circles are two items has no label space between them;
\label{fig:om}
\end{figure}

\begin{example}[Insert]
Figure \ref{fig:om} shows a simple example of the OM data structure.
For simplicity, we choose $N=2^4=16$, so that for items the top-labels $L^t$ are 8-bit integers (above groups as group labels), and the bottom labels are 4-bit integers (below items). 

Figure \ref{fig:om}(a) shows an initial state of the two-level lists and labels. The top-list has head $h^t$ and tail $t^t$ labeled by $0$ and $16^{2} -1$, respectively, and includes four groups $g_1$ to $g_4$ labeled with gap $16$. The bottom-list has head $h_b$ and tail $t_b$ without labels, and includes four items $v_1$ to $v_4$ located in four groups $g_1$ to $g_4$ with same labels $7$. 

Figure \ref{fig:om}(b) shows an intermediate state after a number of \ominsert and \omdelete operations. We can see that there does not exist label space between $v_1$ and $v_2$. Both $v_1$ and $v_2$ are located in the group $g_1$. We get that $g_1$ is full when inserting a new item after $v_1$. 

In Figure \ref{fig:om}(c), a new item $u$ is inserted after $v_1$. But the group $g_1$ is full (no label space after $v_1$), which triggers a \emph{relabel} process. 
That is, the group $g_1$ is split into two groups, $g_1$ and $g$; the old group $g_1$ only has $v_1$;
the newly created group $g$ contains $v_2$ and $v_3$; also, $v_1$ to $v_3$ are assigned $L_b$ with uniform distribution within their own group.  
However, there is no label space between $g_1$ and $g_2$ to insert the new group $g$, which will trigger a \emph{rebalance} operation.
That is, we traverse groups from $g_1$ to $g_4$, where $g_4$ is the first that satisfies $L(g_4) -L(g_1) = 15 > j^2$ ($j =3$). Then, both $g_2$ and $g_3$ are assigned new top-labels as $20$ and $25$, respectively, which have a gap of $5$. Now, $g$ can be inserted after $g_1$ with $L(g) = L(g_2)+ (L(g_2)-L(g_1))/2 = 17$. Finally, we can insert $u$ successfully after $v_1$ in $g_1$, with $L_b(u) = L_b(v_1)+ (15-L_b(v_1))/2 = 11$.
\end{example}

\subsection{Atomic Primitives}

\iffalse
 As shown in Algorithm~\ref{Alg:cas}, the \texttt{CAS} atomic primitive takes three arguments, a variable (location) $x$, an old value $a$ and a new value $b$. It checks the value of the variable $x$, and if it equals to the old value $a$, it updates the pointer to the new value $b$ and then returns \textit{true}; otherwise, it returns \textit{false} to indicate that the updating fails. Here, we use a pair of \emph{angular brackets}, $\langle ... \rangle$, to indicate that the operations in between are executed atomically.
 
\begin{algorithm}[!hbt]
%\small
% \SetKwData{True}{\small TRUE}
% \SetKwData{False}{\small FALSE}
\SetKwData{True}{true}
\SetKwData{False}{false}
\caption{CAS($x, a, b$)}
\label{Alg:cas}
\DontPrintSemicolon
\SetAlgoLined
\SetAlgoNoEnd
    $\langle\,$\lIf{$x = a$}{
        $x \gets b$; \Return \True
    }
    \lElse{
        \Return \False $\rangle$ 
        \textcolor{darkgreen}{\tcc*[f]{$\langle ... \rangle$ atomic}} 
    }
\end{algorithm}

Typically, we assume that the order list $\od$ has at most $2^{32}$ items. In this case, the bottom-labels $L_b$ are 32-bit integers and the top-labels $L^t$ are 64-bit integers.
The modern multicore architectures support atomic primitives for reading or writing 64-bit integers.
\else 

The compare-and-swap atomic primite \texttt{CAS}$(x, a, b)$ atomic primitive takes a variable (location) $x$, an old value $a$ and a new value $b$. It checks the value of $x$, and if it equals $a$, it updates the variable to $b$ and returns \textit{true}; otherwise, it returns \textit{false} to indicate that updating failed. 
%Here, we use a pair of \emph{angular brackets}, $\langle ... \rangle$, to indicate that the operations in between are executed atomically.

\fi %% atomic primitive

\subsection{Lock Implementation}
\iffalse
OpenMP (Open Multi-Processing) \cite{chandra2001parallel} is an application programming interface (API) that supports multi-platform shared-memory multiprocessing programming in C, C++, and Fortran, on many platforms, instruction-set architectures, and operating systems. This paper uses OpenMP (version 4.5) as the threading library to implement the parallel algorithms. 
In this work, the key issue is how to implement the locks for synchronization. 
One solution is to use the OpenMP lock, ``\texttt{omp\_set\_lock}'' and ``\texttt{omp\_unset\_lock}''. 
Each worker will suspend the working task until the specified lock is available. 
The OpenMP lock will be efficient if a lot of work within the locked region.   

The other solution is the spin lock, which can be implemented by the atomic primitive \texttt{CAS}. 
Given a variable $x$ as a lock, the \texttt{CAS} will repeatedly check $x$, and set $x$ from \texttt{false} to \texttt{true} if $x$ is \texttt{false}.
In other words, one worker will busy-wait the lock $x$ until it is released by other workers without suspension.  
The spin lock will be efficient if significantly little work within the locked region exists. 
In this case, suspending has a higher cost than busy waiting for multiple workers. 

%%%%
\begin{algorithm}[!hbt]
%\small
% \SetKwData{True}{\small TRUE}
% \SetKwData{False}{\small FALSE}
\SetKwData{True}{true}
\SetKwData{False}{false}
\SetKwFunction{CAS}{CAS}
\SetKw{Return}{return}
\caption{Lock($x$)}
\label{Alg:lock}
\DontPrintSemicolon
\SetAlgoLined
\SetAlgoNoEnd
    $i \gets 1$\;
    \While{\True}{
        \lIf{$x.\lock = \False \land $\CAS{$x.\lock, \False, \True$}}{\Return}
        $j \gets i$\;
        \lWhile{$j>0$}{$j\gets j - 1$}
        $i \gets 2\times i$\;
    }
\end{algorithm}

Algorithm \ref{Alg:lock} shows an implementation of the spin lock. To reduce the bus traffic, $x.\lock$ is tested before using \texttt{CAS} to set $x.\lock$ from \texttt{false} to \texttt{true} (line 3). Additionally, it is more effective for other workers to \emph{back off} for some duration, giving competing workers a chance to acquire the lock. Typically, especially for our use cases, the large number of unsuccessful tries indicates the longer the worker should back off. Here, we use a simple strategy that exponentially increases the back off time for each try (lines 1 and 4 - 6), where $i$ and $j$ are local variables without increasing the bus traffic~\cite{herlihy2020art}.    

\else

OpenMP (Open Multi-Processing) \cite{chandra2001parallel} is an application programming interface (API) that supports multi-platform shared-memory multiprocessing programming in C, C++, and Fortran, on many platforms, instruction-set architectures, and operating systems. This paper uses OpenMP (version 4.5) as the threading library to implement the parallel algorithms. 
In this work, the key issue is how to implement the locks for synchronization. 
One solution is to use the OpenMP lock, ``\texttt{omp\_set\_lock}'' and ``\texttt{omp\_unset\_lock}''. 
Each worker will suspend the working task until the specified lock is available. 
The OpenMP lock will be efficient if a lot of work within the locked region.   

The other solution is the spin lock, which can be implemented by the atomic primitive \texttt{CAS}. 
Given a variable $x$ as a lock, the \texttt{CAS} will repeatedly check $x$, and set $x$ from \texttt{false} to \texttt{true} if $x$ is \texttt{false}.
In other words, one worker will busy-wait the lock $x$ until it is released by other workers without suspension.
The spin lock will be efficient if very little work within the locked region is needed. 
In this case, suspending has a higher cost than busy waiting for multiple workers. 
Typically, especially for our use cases, a large number of unsuccessful tries indicates that workers should back off longer before trying again. We exponentially increases the \emph{back off} time for each try~\cite{herlihy2020art}.   

\fi %%%% lock implementation

\section{Parallel Order Maintenance Data Structure}
%We observe that most of the $k$-order operation is to compare the order of two items. For this, without locking any items, our \texttt{Parallel-OMOrder}$(\od, x,y)$ operation that can determine the order of $x$ and $y$ since the  Invariant in Definition \ref{def:order} is hold  \texttt{Parallel-OMInsert}$(od, x, y)$ operatio

In this section, we present the parallel version of the OM data structure. 
We start from the parallel \texttt{Delete} operations. Then, we discuss the parallel \texttt{Insert} operation and show that the Order Snapshot is preserved at any steps including the relable process, which is the main contribution of this work.
%Both the parallel \texttt{Delete} and \texttt{Insert} use locks for synchronization. 
Finally, we present the parallel \texttt{Order} operation, which is lock-free and thus can be executed highly in parallel. 

\subsection{Parallel Delete}
\subsubsection{Algorithm}
Algorithm \ref{alg:delete} shows the detail steps of parallel \texttt{Delete}. 
For each item $x$ in $\od$, we use a boolean status $x.\live$ to indicate if $x$ is in $\od$ or has been removed. Initially $x.\live$ is \texttt{true}.  
We use the atomic primitive \texttt{CAS} to set $x.\live$ from \texttt{true} to \texttt{false}, and a repeated delete of $x$ will return \texttt{fail} (line 1). In lines 2 - 6 and 13, we remove $x$ from the bottom-list in $\od$. 
To do this, we first lock $y = x.\pre$, $x$, and $x.\Next$ in order to avoid deadlock (lines 2 - 4). 
Here, after locking $y$, we have to check that $y$ still equals $x.\pre$ in case $x.\pre$ is changed by other workers (line 3). Then, we can safely delete $x$ from the the bottom-list, and set $x$'s $\pre$, $\Next$, $L_b$, and $\group$ to empty (line~6). Finally, we unlock all locked items in reverse order (line~13). 
Finally, for the group $g = x.\group$, we need to delete $g$ when it is empty, which is analogous to deleting $x$ (lines 7 - 12). 

\begin{algorithm}[!htb]
%\footnotesize
%\small
\SetKwRepeat{Do}{do}{while}
\SetAlgoNoEnd
%%%% here use order is better, since order : k-order or topological order in graph
\caption{Parallel-Delete($\od, x$)}
\label{alg:delete}
\DontPrintSemicolon
\SetKwInOut{Input}{input}\SetKwInOut{Output}{output}
\SetKwFunction{Relabel}{Relabel}
\SetKwFunction{TempLabel}{Temp-Label}
\SetKwFunction{group-label}{group-label}
\SetKwFunction{Lock}{Lock}
\SetKwFunction{Unlock}{Unlock}
\SetKwFunction{CAS}{CAS}
\SetKwRepeat{Do}{do}{while}
\SetKw{Return}{return}
\SetKw{Not}{not}
\SetKw{Goto}{goto}
\SetKw{With}{with}
\SetKwData{True}{true}
\SetKwData{False}{false}
\SetKwData{Fail}{fail}

\lIf{\Not \CAS{$x.\live, \True, \False$}}{\Return \Fail} 
$y \gets x.\pre$; 
%\lIf{$y = \varnothing$}{\Return \tcp*[!h]{$x$ has been deleted}} 
\Lock{$y$}\; %%%% after lock y (if x.pre is y), x can not be deleted.
\lIf{$y\neq x.\pre$}{{\Unlock{$y$}; \Goto line 1}} 
\Lock{$x$}; \Lock{$x.\Next$}\; $g\gets x.\group$\;
delete $x$ from bottom-list; set $x.\pre$, $x.\Next$, $L_b(x)$, and $x.\group$ to $\varnothing$

\If{$|g|=0~\land~$\CAS{$g.\live, \True, \False$}}{
    $g'\gets g.\pre$ \label{alg:delete-g};
    %\lIf{$g' = \varnothing$} {\Goto line \ref{alg:delete-end} \tcp*[!h]{$g$ has been deleted}}
    \Lock{$g'$}\; 
    \lIf{$g' \neq g.\pre$}{\Unlock{$g'$}; \Goto line \ref{alg:delete-g}} \Lock{$g$}; \Lock{$g.\Next$}\;
    delete $g$ from top-list; set $g.\pre$, $g.\Next$, and $L(g)$ to $\varnothing$\;
    \Unlock{$g.\Next$}; \Unlock{$g$}; \Unlock{$g'$} 
} 
\Unlock{$x.\Next$}; \Unlock{$x$}; \Unlock{$y$}; \label{alg:delete-end}
\end{algorithm} %%%% InsertEdges

We \emph{logically} delete items by setting their flags (line 1). One benefit of such a method is that we can delay the \emph{physical} deleting of items (lines 2 - 13). 
The physical deleting can be batched and performed lazily at a convenient time, reducing the overhead of synchronization. Typically, this technique is used in linked lists for delete operations~\cite{herlihy2020art}. In his work, we will not test the delayed physical delete operations.  

Obviously, during parallel \texttt{Delete} operations, the labels of other items are not affected and the Order Snapshot is maintained. 

\subsubsection{Correctness}
For deleting $x$, we always lock three items, $x.\pre$, $x$, and $x.\Next$ in order. Therefore, there are no blocking cycles, and thus deletion is deadlock-free. 

\subsubsection{Complexities}
Suppose there are $m$ items to delete in the OM data structure. The total work is $O(m)$. In the best case, $m$ items can be deleted in parallel by $\mathcal P$ workers with $O(1)$ depth, so that the total running time is $O(m/\mathcal P)$. In the worst-case, $m$ items have to be deleted one by one, e.g. $\mathcal P$ workers are blocked as a chain, with $O(m)$ depth, so that the total running time is $O(m/\mathcal P + m)$. 

However, when deleting multiple items in parallel, a blocking chain is unlikely to appear, and thus the worst-case has a low probability to happen.

\subsection{Parallel Insert}
\subsubsection{Algorithm}
Algorithm \ref{alg:om-insert} shows the detailed steps for inserting $y$ after $x$. 
Within this operation, we lock $x$ and its successor $z = x.\Next$ in that order (lines 1 and 7).
For obtaining a new bottom-label for $y$, if $x$ and $z$ is in the same group, $L_b(z)$ is the right bound; otherwise, $N$ is the right bound supposing $L_b$ is a $(\log N)$-bit integer (line 2). 
If there does not exist a label gap in the bottom-list between $x$ and $x.\Next$, we know that $x.\group$ is full, and thus the \texttt{Relabel} procedure is triggered to make label space for $y$ (line 3). 
Then, $y$ is inserted into the bottom-list between $x$ and $x.\Next$ (line 6), in the same group as $x$ (line 4), by assigning a new bottom-label (line 5). 

The \texttt{Relabel}$(x)$ procedure splits the full group of $x$. 
We lock $x$'s group $g_0$ and $g_0$'s successor $g.\Next$ (line 9). We also lock all items $y\in g_0$ except $x$, as $x$ is already locked in line 1 (line 10). 
To split the group $g_0$ into multiple new smaller groups, we traverse the items $y \in g_0$ in reverse order by three steps (lines 11 - 15). 
First, if there does not exist a label gap in the top-list between $g_0$ and $g_0.\Next$, the \texttt{Rebalance} procedure is triggered to make label space for inserting a new group with assigned labels (lines 12 and 13). 
Second, we split $\frac{log N}{2}$  items $y$ from $g_0$ in reverse order to the new group $g$, which maintains the Order Snapshot (line 14). 
Third, we assign new $L_b$ to all items in the new group~$g$ by using the \texttt{AssignLabel} procedure (line 15), which also maintains the Order Snapshot. 
The for-loop (lines 11 - 15) stops if less than $\frac{log N}{2}$ items are left in $g_0$. We assign new $L_b$ to all left items in $g_0$ by using the \texttt{AssignLabel} procedure (line 16). 
Finally, we unlock all locked groups and items (line 17). 

In the \texttt{Rebalance}$(g)$ procedure, we make label space after~$g$ to insert new groups. Starting from $g.\Next$, we traverse groups $g'$ in order until $w > j^2$ by locking $g'$ if necessary ($g$ and $g.\Next$ are already locked in line 9), where $j$ is the number of visited groups and $w$ is the label gap $L(g') - L(g)$ (lines 19 - 22). 
That means $j$ items will totally share $w>j^2$ label space. All groups whose labels should be updated are added to the set $A$ (line 21). We assign new labels to all groups in $A$ by using the \texttt{AssignLabel} procedure (line 23), which maintains the Order Snapshot.  
Finally, we unlock groups locked in line 21 (line 24). 

Notably, in the \texttt{AssignLabel}$(A, \mathcal L, l_0, w)$ procedure, we assign labels without affecting the Order Snapshot, where the set $A$ includes all elements whose labels need updating, $\mathcal L$ is the label function, $l_0$ is the starting label, and $w$ is the label space.
Note that, $\mathcal L$ can correctly return the bounded labels, e.g., $L_b(x.\Next) = N$ when $x$ is at the tail of its group $x.\group$. 
For each $z \in A$ in order, we first correctly assign a temporary label $\overline{\mathcal L}(z)$ (line 27), which can replace its real label $\mathcal L(z)$ at the right time by using stack $S$ (lines 28 - 32). 
Specifically, for each $z\in A$ in order, if its temporary label $\overline{\mathcal L}(z)$ is between $\mathcal L(z.pre) $ and $ \mathcal L(z.\Next)$, we can safely \emph{replace} its label by updating ${\mathcal L}(z)$ as $\overline{\mathcal L}(z)$ (lines 29 and 30), which maintains the Order Snapshot; otherwise, $z$ is added to the stack $S$ for further propagation (line 32). 
For the propagation, when one element $z$ replaces the labels (line 30), which means all elements in stack $S$ can find enough label space, each $x\in S$ can be popped out by replacing its label (line 31). This propagation still maintains the Order Snapshot.

%%%%%%%%%%%%%%%%%%%%%%%%%%%%%%%%%%
%% one direction rebalances 
%%% add this to the complexities analysis. 
%%%%%%%%%%%%%%%%%%%%%%%%%%%%%%%

% \paragraph{}
% Note that, 

%%%%
\begin{algorithm}[!htb]
%\small
%\footnotesize
\SetKwRepeat{Do}{do}{while}

\DontPrintSemicolon
\SetAlgoNoEnd
\SetAlgoLined
%%%% here use order is better, since order : k-order or topological order in graph
\caption{Parellel-Insert($\od, x, y$)}
\label{alg:om-insert}

\SetKwInOut{Input}{input}\SetKwInOut{Output}{output}
\SetKwFunction{Relabel}{Relabel}
\SetKwFunction{AssignLabel}{AssignLabel}
\SetKwFunction{Rebalance}{Rebalance}
\SetKwFunction{TempLabel}{Temp-Label}
\SetKwFunction{group-label}{group-label}
\SetKwFunction{Lock}{Lock}
\SetKwFunction{Unlock}{Unlock}
\SetKw{With}{$:$}
\SetKw{In}{in}
\SetKwData{True}{true}
\SetKwRepeat{Do}{do}{while}

%% sublabel is not changed after assigned. 
\Lock{$x$}; $z\gets x.\Next$; \Lock{$z$}\;
\leIf{$x.\group = z.\group$}{$b\gets L_b(z)$}{$b \gets N$}
\lIf{$b - L_b(x) < 2$}{
    %$\version\texttt{++}$; 
    \Relabel{$x$}; 
    %$\version\texttt{++}$
}
%$L^t(y.\group) = L^t(x.\group)$\; 
insert $y$ into bottom-list between $x$ and $x.\Next$\;
$L_b(y) \gets L_b(x) + \lfloor (b - L_b(x)) / 2 \rfloor $\;
$y.\group \gets x.\group$\;
\Unlock{$x$}; \Unlock{$z$}\;

\medskip
\SetKwProg{myproc}{procedure}{ \Relabel{$x$}}{}
\myproc{}{
   $g_0 \gets x.\group$; \Lock{$g_0$}; \Lock{$g_0.\Next$};
  
  \Lock all items $y \in g_0$ with $y \neq x$ in order\;
  %\For{$y \in g_0$ \rm{in reverse order until less than ${(\log N)}/{2}$ items left in $g_0$}}{
   \For{$y \in g_0$ \rm{in reverse order until less than $\frac{\log N}{2}$ items left in $g_0$}}{  
        \lIf{$L(g_0.\Next) - L(g_0) < 2 $}{\Rebalance{$g_0$}}
        insert a new group $g$ into the top-list after $g_0$ with $L(g) = (L(g_0.\Next) - L(g_0))/2$\; 
        split out $\frac{\log N}{2}$ items $y$ into $g$\;
        \AssignLabel{$g, L_b, 0, N$}
  }
  \AssignLabel{$g_0, L_b, 0, N$}\; 
  \Unlock $g_0.\Next$, $g_0$, and all items $y \in g_0$ with $y \neq x$\; 
}
    
\medskip
\SetKwProg{myproc}{procedure}{ \Rebalance{$g$}}{}
\myproc{}{
    $g' \gets g.next$;  $j \gets 1$; 
    $w \gets L(g') - L(g)$; $A \gets \emptyset$\;
    %$n \gets $ the number of new groups inserted\;
    %$n = $
    \While{$w \leq j^2$}{ %\lor j < n
        $A \gets A \cup \{g'\}$; $g' \gets g'\!.\Next$; \Lock{$g'$} \label{alg:lock-group}\; 
        $j \gets j + 1$;
        $w \gets L(g') - L(g)$\;
         
    }
    \AssignLabel{$A, L^t, L^t(g), w$}\;
    \Unlock all locked groups in line \ref{alg:lock-group}.
}

\medskip
\SetKwProg{myproc}{procedure}{ \AssignLabel{$A, \mathcal L, l_0, w$}}{}
\myproc{}{
    %\For{$k\gets1$ \KwTo $i$ \KwBy $1$}{
    %    ${L^t'(g')} = L^t(g) + w_j/j$
    %}
    %$k \gets 1$\; %: \normalfont{in order with $k$ as index}
    $S \gets $ empty stack; $k \gets 1$; $j\gets |A|+1$ \;
    \lFor{$z \in A~\mathrm{in~order}$}{$\overline{\mathcal L}(z) = l + k \cdot w /j$; $k\gets k + 1$}
    \For{$z \in A~\mathrm{in~order}$ }{
        \If{$\mathcal L(z.pre) < \overline{\mathcal L}(z) < \mathcal L(z.\Next)$}
        %\If{$z.\pre \preceq z \preceq z.\Next$ \rm is \True by using $\overline{\mathcal L(z)}$}
        {
            $\mathcal L(z) \gets \overline{\mathcal L}(z)$\;
            \lWhile{$S \neq \emptyset$}{
                $x \gets S.\Pop()$; $\mathcal L(x) \gets \overline{\mathcal L}(x)$
            }
        }
        \lElse{
            $S.\Push(z)$
        }
    }
    
}
\end{algorithm} %%%% InsertEdges

\iffalse 
\mytodo{
\paragraph{}
Note that, 
}
\fi 

\begin{example}[Parallel Insert]
Figure \ref{fig:om} shows an example for parallel \texttt{Insert}. 
In Figure \ref{fig:om}(b), we lock $v_1$ and $v_2$ in order when inserting $u$ after $v_1$. However, there is no label space, and the group $g_1$ is full, which triggers the \texttt{Relabel} procedure. For the first step of relabelling, the other item $v_3$ in group $g_1$ is locked to split the group $g_1$.  

In Figure \ref{fig:om}(c), we lock $g_1$ and $g_2$ in order when inserting a new group $g$ after $g_1$, which triggers the \texttt{Rebalance} procedure on the top-list. For rebalancing, $g_3$ and $g_4$ are locked in order. 
The new temporary labels $\overline{L^t}$ of $g_2$ and $g_3$ are generated as $\overline{20}$ and $\overline{25}$. 
To replace real labels with temporary ones, we traverse $g_2$ and $g_3$ in order. 
First, we find that $L(g_1)< \overline{L}(g_2) < L(g_3)$ as $15 < \overline{20} < 17$ is false, so that $g_2$ is added to the stack such that $S=\{g_2\}$. 
Second, when traversing $g_3$, we find that $L(g_2)< \overline{L}(g_3) < L(g_4)$ as $16 < \overline{25} < 30$ is true, so that $L(g_3)$ is replaced as $25$. In this case, the propagation of $S$ begins and $g_2$ is popped out with $L(g_2)$ replaced as $20$.
Finally, $g_3$ and $g_4$ are unlocked and the \texttt{Rebalance} procedure finishes.

%%%% v2 and v3 should have large labels to decrease. 
After rebalancing, the new group $g$ can be insert after $g_1$ with $L(g_1) = 17$. 
Relabeling continues. The item $v_3$ is spitted out to $g$ with $L_b(v_3) = 15 \land L^t(v_3) = 17$ maintaining the Order Snapshot; similarly, $v_2$ is also spitted out to $g$. Now, both $v_2$ and $v_3$ require new $L_b$ to be assigned by the \texttt{AssignLabel} procedure. 
The new temporary label $\overline{L_b}$ of $v_2$ and $v_3$ are generated as $\overline{5}$ and $\overline{10}$, respectively. 
For replacing, we traverse $v_2$ and $v_3$ in order by two steps. 
First, for $v_2$, we find that $\overline{L_b}(v_2) < L_b(v_3)$ is true, so that $L_b(v_2)$ is replaced as $5$.
Second, for $v_3$, we find that $L_b(v_2)< \overline{L_b}(v_3)$ is true, so that $L_b(v_3)$ is replaced as $10$
There is no further propagation since the stack $S$ is empty. 
Now, only one item $v_1$ is left in $g_1$ and $L_b(v_1)$ is set to $7$. Finally, the new item $u$ is inserted after $v_1$ in $g_1$ with $L_b(u) = 11 \land L^t(u) = 15$.
\end{example}

\begin{figure}[!htb]
\centering
\includegraphics[width=0.7\linewidth]{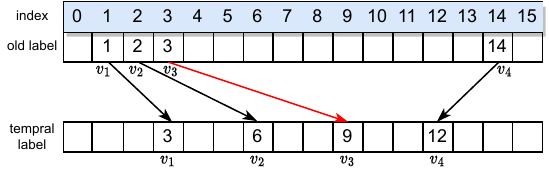} %  [width=\linewidth]
\caption{An example of the \texttt{AssignLabel} procedure.}
\label{fig:assign}
\end{figure}

\begin{example}[Assign Label]
In Figure \ref{fig:assign}, we show an example how the \texttt{AssignLabel} procedure preserves the Order Snapshot. The label space is from 0 to 15, shown as indices. There are four items $v_1$, $v_2$, $v_3$, and $v_4$ with initial labels $1$, $2$, $3$, and $14$, respectively; also, four temporary labels, $3, 6, 9$, and $12$, are assigned with uniform distribution to them. 
We traverse items from $v_1$ to $v_4$ in order. 
First, $v_1$ and $v2$ are added to the stack $S$. Then, $v_3$ can safely replace its old label with its new temporary label $9$, which makes space for $v2$ that is at the top of $S$. So, we pop out $v_2$ from $S$ and $v_2$ get its new label $6$, which makes space for $v_1$ that is at the top of $S$. So, we pop out $v_1$ from $S$ and $v_1$ gets its new label $3$. 
Finally, $v_4$ can safely get its new label $12$. 
In a word, updating the $v_3$'s label will repeatedly make space for $v_2$ and $v_1$ in the stack. 
During such a process, we observe that each time an old label is updated with a new temporal label, the labels always correctly indicate the order. Therefore, the Order Snapshot is always preserved and parallel \texttt{Order} operations can take place.
\end{example}

\subsubsection{Correctness}
All items and groups are locked in order. Therefore, there are no blocking cycles, and thus parallel insertions and deletions are deadlock-free.

We prove that the Order Snapshot is preserved during parallel \texttt{Insert} operations. In Algorithm \ref{alg:om-insert}, there are two cases where the labels are updated, splitting groups (lines 11 - 15) and assigning labels (lines 15, 16, and 23) by using the \texttt{AssignLabel} procedure (lines 25 - 32). 

\begin{theorem} 
 When splitting full groups (line 14), the Order Snapshot is preserved.  
 \label{th:split-group}
\end{theorem}
\begin{proof}
The algorithm splits $\frac{\log N}{2}$ items $y$ out from $g_0$ into the new group $g$ (line 14), where each $y\in g_0$ is traversed in reverse order within the for-loop (lines 11 - 15). For this, the invariant of the for-loop is that $y$ has largest $L_b$ within $g_0$; the new group $g$ has $L(g) > L(g_0)$; also, $y$ satisfies the Order Snapshot: 
\begin{equation*} \begin{split}
&(\forall x \in g_0: x \neq y \implies L_b(y) > L_b(x)) \\
&~\land~(L(g_0) < L(g))~\land~(y.\pre \preceq y \preceq y.\Next)
\end{split} \end{equation*}

% \begin{equation*} 
% (\forall x \in g_0: x \neq y \implies L_b(y) > L_b(x)) 
% ~\land~(L(g_0) < L(g))~\land~(y.\pre \preceq y \preceq y.\Next)
% \end{equation*}
We now argue the for-loop preserve this invariant: 
\begin{itemize} %[leftmargin=*]
\item[--] $\forall x \in g_0: x \neq y \implies L_b(y) > L_b(x)$ is preserved as $y$ is traversed in reserve order within $g_0$ and all other items $y'$ that have $L_b(y) < L_b(y')$ are already spitted out from $g_0$. 

\item[--] $L(g_0) < L(g)$ is preserved as $g$ is new inserted into top-list after $g_0$. 

\item[--] $y.\pre \preceq y$ is preserved as we have $y\in g \land y.\pre \in g_0$ and $L(g_0) < L(g)$. 

\item[--] $y \preceq y.\Next$ is preserved as if $y$ and $y.\Next$ all in the same group $g$, we have $L_b(y) < L_b(y.\Next)$; also, if $y$ and $y.\Next$ in different groups, we have $y$ is the first item moved to $g$ or $y$ is still located in $g_0$,  which their groups indicates the correct order.   
\end{itemize}

At the termination of the for-loop, the group $g$ is split into multiple groups preserving the Order Snapshot. 
\end{proof}

\begin{theorem} 
When assigning labels by using the \texttt{AssignLabel} procedure (lines 25 - 32), the Order Snapshot is preserved.  
 \label{th:assign-label}
\end{theorem}
\begin{proof}
The \texttt{AssignLabel} procedure (lines 25 - 32) assigns labels for all items $z \in A$. The temporal labels are first generated in advance (line 27). Then, the for-loop replaces the old label with new temporal labels (lines 28 - 32). The key issue is to argue the correctness of the inner while-loop (line 31).
%Before the proof, we define $z$'s Order Snapshot as $z.\pre \preceq \overline{z} \preceq z.\Next$ when $z$ uses the temporal label $\mathcal L(z)$ 
%The invariant of this-loop is that all items $z' \in S \land z' \neq S.top$ not satisfy the Order Snapshot when $z'$ uses temporal label; 
The invariant of this inner while-loop is that the top item in $S$ has a temporal label that satisfies the Order Snapshot:
\begin{equation*} \begin{split}
%& \mathcal L(z.\pre) < \mathcal L(z) < \mathcal L(z.\Next)  \\
%&~\land~(z' = S.\TOP \implies \mathcal L(z'.\pre) < \overline{\mathcal L}(z') < \mathcal L(z'.\Next)) 
& (\forall y \in S: (y \neq S.\TOP \implies y \preceq S.\TOP) ~\land~ y \preceq z) \\
&~\land~x = S.\TOP \implies \mathcal L(x.\pre) < \overline{\mathcal L}(x) < \mathcal L(x.\Next) 
\end{split} \end{equation*}
%The invariant initially holds as $\mathcal L(z)$ is correctly replaced by temporal label $\overline{\mathcal L(z)}$ only when such temporal label satisfy the Order Snapshot (line 30). 
The invariant initially holds as $\mathcal L(z)$ is correctly replaced by the temporal label $\overline {\mathcal L}(z)$ in line 30 and $z$ is $x.\Next$, so that $\overline{\mathcal L}(x) < \mathcal L(z)$; also, we have $\mathcal(x.\pre)< \overline{\mathcal L}(x)$ as if it is not satisfied, $x$ should not been added into $S$, which causes contratiction. 
We now argue the while-loop (line 31) preserves this invariant: 
\begin{itemize} % [leftmargin=*]
%\item[--] $ \mathcal L(z.\pre) < \mathcal L(z) < \mathcal L(z.\Next) $ is preserved as $\mathcal L(z)$ is correctly replaced by temporal label $\overline{\mathcal L(z)}$ (line 31). 
\item[--] $\forall y \in S: (y \neq S.\TOP \implies y \preceq S.\TOP) ~\land~ y \preceq z$ is preserved as all items in $S$ are added in order, so the top item always has the largest order; also, since all item in $A$ are traversed in order, so $z$ has the larger order than all item in $S$. 

\item[--] $x = S.\TOP \implies  \overline{\mathcal L}(x) < \mathcal L(x.\Next)$ is preserved as $\mathcal L(x.\Next)$ is already replace by the temporal label $\overline {\mathcal L}(x.\Next)$ and $x$ is precede $x.\Next$ by using temporal labels. 

\item[--] $x = S.\TOP \implies \mathcal L(x.\pre) < \overline{\mathcal L}(x)$ is preserved as if such invariant is not satisfied, $x$ should not been added into $S$ and can be safely replace the label by its temporal label, which causes contradiction. 
\end{itemize}

At the termination of the inner while-loop, we get $S=\emptyset$, so that all items that precede $z$ have replaced new labels maintaining the Order Snapshot. At the termination of the for-loop (lines 28 -32), all items in $A$ have been replaced with new labels. 
\end{proof}

\subsubsection{Complexities}
For the sequential version, it is proven that the amortized time is $O(1)$. The parallel version has some refinement. That is, the \texttt{AssignLabel} procedure traverses the locked items two times for generating temporary labels and replacing the labels, which cost amortized time $O(1)$. Thus, if $m$ items are inserted in parallel, the total amortized work is $O(m)$. 
In the best case, $m$ items can be inserted in parallel by $\mathcal P$ workers with amortized depth $O(1)$, so that the amortized running time is $O(m/\mathcal P)$. 

The worst-case can easily happen when all insertions accrued in the same position of $\od$. The relabel procedure is triggered with the constant amortized work $\mathcal W = O(1)$ for each inserted item.
In the worst-case, $m$ items have to be inserted one-by-by, e.g. $\mathcal P$ workers simultaneously insert items at the head of $\od$ with amortized depth $O(m)$, and thus the amortized running time is $O(m/\mathcal P + m)$.

Such worst-case can be improved by batch insertion. The idea is that we first allocate enough label space for $m/\mathcal P$ items per worker, then $\mathcal P$ workers can insert items in parallel. However, this simple strategy requires pre-processing of $\od$ and does not change the worst-case time complexity.   

%Such a worst-case is easy to happen with a high probability. One strategy is that  

%\mytodo{relabel procedure complexities?}

\subsection{Parallel Order}
\subsubsection{Algorithm}
Algorithm \ref{alg:om-order} shows the detailed steps of \texttt{Order}.
When comparing the order of $x$ and $y$, they must not have been be deleted (line 1).
We first compare the top-labels of $x$ and $y$ (lines 2 - 5). Two variables, $t$ and $t'$, obtain the values of $L^t(x)$ and $L^t(y)$ for comparison (line 2), and the result is stored as $r$. After that, we have to check $L^t(x)$ or $L^t(y)$ has been updated or not; if that is the case, we have to redo the whole procedure (line 5). 
Second, we compare the bottom-labels of $x$ and $y$, if their top-lables are equal (lines 6 - 9). Similarly, two variables, $b$ and $b'$, obtain the value of $L_b(x)$ and $L_b(t)$ for comparison (line 7), and the result is stored as $r$. After that, we have to check whether four labels are updated or not; if anyone label is the case, we have to \emph{{redo}} the whole procedure (lines 8 and 9). We can see our parallel \texttt{Order} is lock-free so that it can execute highly in parallel. 
During the order comparison, $x$ or $y$ can not be deleted (line 10).
We return the result at line 11.

%%%%%%%%%%%%%%%%%%%%%%%%%
% the ABA problem. 
% There are no ABA problems, prove that. 
% Argue it in correctness with different cases. The condition in line 8 will not detect the inconsistency. Is it a problem?
%%%%%%%%%%%%%%%%%%%%%%%%%%%%

%%%%
\begin{algorithm}[!htb]
%\footnotesize
%\small
\SetKwRepeat{Do}{do}{while}
\SetAlgoNoEnd
%%%% here use order is better, since order : k-order or topological order in graph
\caption{Parallel-Order($\od, x, y$)}
\label{alg:om-order}
\DontPrintSemicolon
\SetKwInOut{Input}{input}\SetKwInOut{Output}{output}
\SetKwFunction{Relabel}{Relabel}
\SetKwFunction{TempLabel}{Temp-Label}
\SetKwFunction{group-label}{group-label}
\SetKwFunction{Lock}{lock}
\SetKwFunction{Unlock}{unlock}
\SetKwRepeat{Do}{do}{while}
\SetKw{Return}{return}
\SetKw{Goto}{goto}
\SetKwData{True}{true}
\SetKwData{False}{false}
\SetKwData{Fail}{fail}
%await $L_b(x) \neq L^t(x) \neq  L_b(y) \neq L^t(y) \neq \varnothing$\;
\lIf{$x.\live = \False \lor y.\live = \False$ }{\Return \Fail \label{alg:om-order-first-line}}
$t, t', r \gets L^t(x), L^t(y), \varnothing$\; 
\If{$t\neq t'$}{
    $r \gets t < t'$\;
    \lIf{$t \neq L^t(x) \lor t' \neq L^t(y)$}{ \Goto line \ref{alg:om-order-first-line}}
}
\Else{
    $b, b' \gets L_b(x), L_b(y)$;
    $r \gets b < b'$\;
    %\newline \makebox[1cm]{}
    \If{$t \neq L^t(x) \lor t' \neq L^t(y)  \lor b \neq L_b(x) \lor b' \neq L_b(y)$}{ \Goto line \ref{alg:om-order-first-line}}
}
\lIf{$x.\live = \False \lor y.\live = \False$ }{\Return \Fail}
\Return $r$\;
% \lIf{$L^t(x) < L^t(y)$}{ 
%     \Return \True
% }
% \lElseIf{$L^t(x) = L^t(y)~\land~L_b(x) < L_b(y)$}{
%     \Return \True
% }
% \Return \False
\end{algorithm} %%%% InsertEdges

It is true that there is an \emph{ABA problem}. That is, $L^t(x)$ and $L^t(y)$ are possibly updated multiple times but remain the same values as the $t$ and $t'$ (line 5). In other words, $L^t(x)$ and $L^t(y)$ are updated but may not be identified when comparing $t$ and $t'$ (line 4), which may lead to a wrong result. Also, line 8 has the same problem.  
To solve this problem, each top-label or bottom-label, $L^t$ or $L_b$, includes an $8$-bit counter to record the version. Each time, the counter increases by one once its corresponding label is updated. With this implementation, we can safely check whether the label is updated or not merely by comparing the values (lines 5 and 8).

\begin{example}[Order]
In Figure \ref{fig:assign}, we show an example to determine the order of $v_2$ and $v_3$ by comparing their labels. Initially, both $v_2$ and $v_3$ have old labels, $2$ and $3$. After the \texttt{Relabel} procedure is triggered, both $v_2$ and $v_3$ have new labels, $6$ and $9$, in which the Order Snapshot is preserved. However, it is possible that \texttt{Relabel} procedures are triggered in parallel. 
We first get $\mathcal L(v_3) = 3$ (old label) and second get $\mathcal L(v_2) = 6$ (new label), but it is incorrect for $\mathcal L(v_2) > \mathcal L(v_3)$. 
After we get $\mathcal L(v_2) = 6$, the value of $\mathcal L(v_2)$ has to be already updated to $9$ since the Order Snapshot is maintained. In this case, we redo the whole process until $\mathcal L(v_2)$ and $\mathcal L(v_3)$ are not updated during comparison. Thus, we can get the correct result of $\mathcal L(v_2) < \mathcal L(v_3)$ even the relabel procedure is executed in parallel.  
\end{example}

\subsubsection{Correctness}
We have proven that parallel \texttt{Insert} preserves the Order Snapshot even though relabel procedures are triggered, by which labels correctly indicate the order. In this case, it is safe to determine the order for $x$ and $y$ in parallel. 
We first argue the top-labels (lines 2 - 5). 
The problem is that we first get $t \gets  L^t(x)$ and second get $t' \gets  L^t(y)$ successively (line 2), by which $l$ and $l'$ may be inconsistent, due to a \texttt{Relabel} procedure may be triggered. 
To argue the consistency of labels, there are two cases:
1) both $t$ and $t'$ obtain old labels or new labels, which can correctly indicate the order;
2) the $t$ first obtains an old label and $t'$ second obtains a new label, which may not correctly indicate the order as $x$ may already update with a new label, and vice versa; if that is the case, we redo the whole process.

    %\item The $t$ first obtain a new label and $t'$ second obtain an old label, which may not correctly indicate the order, as $y$ may already update with new labels. If that is the case, we redo the whole process.

On the termination of parallel \texttt{Order}, the invariant is that $t$ and $t'$ are consistent and thus correctly indicate the order.
The bottom-labels are analogous (lines 6 - 9). %%  so we omit the proof here.

\subsubsection{Complexities}
For the sequential version, the running time is $O(1)$. For the parallel version, we have to consider the frequency of redo. 
It has a significantly low probability that the redo will be triggered. This is because the labels are changed by the \texttt{Relabel} procedure, which is triggered when inserting $\Omega(\log N)$ items. Even if the labels of $x$ and $y$ are updated when comparing their order, it still has a tiny probability that such label updating happens during the comparison of labels (lines 4 and 7).   

Thus, supposing $m$ items are comparing orders in parallel, the total work is $O(m)$, and the depth is $O(1)$ with a high probability.
So that the running time is $O(m/\mathcal P)$ with high probability.

\section{Experiments}
We report on experimental studies for our three parallel order maintenance operations, \texttt{Order}, \texttt{Insert}, and \texttt{Delete}. We generate four different test cases to evaluate their parallelized performance. All the source code is available on GitHub\footnote{\url{https://github.com/Itisben/Parallel-OM.git}}.  

%Unfortunately, we were unable to implement the parallel OM data structure in~\cite{utterback2016provably}, which is combine with series-parallel maintenance and has totally different mechanism with our implementation. Thus, we do not experimentally compare such approach with ours. 

\subsection{Experiment Setup}
The experiments are performed on a server with an AMD CPU (64 cores, 128 hyperthreads, 256 MB of last-level shared cache) and 256 GB of main memory. The server runs the Ubuntu Linux (22.04) operating system. All tested algorithms are implemented in
C++ and compiled with g++ version 11.2.0 with the -O3 option. OpenMP \footnote{\url{https://www.openmp.org/}} version 4.5 is used as the threading library. 
We choose the number of workers exponentially increasing as $1, 2, 4, 8, 16, 32$ and $64$ to evaluate the parallelism. 
With different numbers of workers, we perform every experiment at least 100 times and calculate the mean with 95\% confidence intervals.

In our experiment, for easy implementation, we choose $N=2^{32}$, a $32$-bit integer, as the capacity of $\od$. In this case, the bottom-lables $L_b$ are $32$-bit integers, and the top-lables $L^t$ are 64-bit integers. 
One advantage is that reading and writing such $32$-bit or $64$-bit integers are atomic operations in modern machines. 
There are initially 10 million items in the order list $\od$. To test our parallel OM data structure, we do four experiments:
\begin{itemize} % [leftmargin=*,noitemsep] %[noitemsep,topsep=0pt,leftmargin=*]
    \item[--] \emph{Insert}: we insert 10 million items into $\od$. 
    
    \item[--] \emph{Order}: for each inserted item, we compare its order with its successive item, so that it has 10 million \texttt{Order} operations.
    
    \item[--] \emph{Delete}: we delete all inserted items, a total of 10 million times.
    
    \item[--] \emph{Mixed}: again, we insert 10 million items, mixed with 100 million \texttt{Order} operations. For each inserted item, we compare its order with its ten successive items, a total of 100 million times order comparison. 
    This experiment is to test how often ``redo'' occurs in the \texttt{Order} operations when there are parallel \texttt{Insert} operations.
    The reason for this experiment is that many \texttt{Order} operations are mixed with few \texttt{Insert} and \texttt{Delete} operations in real applications, as shown in Figure \ref{fig:om-core-maint}.    
\end{itemize}

For each experiment, we have four test cases by choosing different numbers of positions for inserting:  
\begin{itemize}%[leftmargin=*,noitemsep] % [noitemsep,topsep=0pt,leftmargin=*]
    \item[--] \emph{No Relabel} case: we have 10 million positions, the total number of initial items in $\od$, so that each position averagely has \textbf{$1$} inserted item. Thus, it almost has {\emph{no}} \texttt{Relabel} procedures triggered when inserting. 
    
    \item[--] \emph{Few Relabel} case: we randomly choose 1 million positions from 10 million items in $\od$, so that each position averagely has {10} inserted items. Thus, it is possible that a {\emph{few}} \texttt{Relabel} procedures are triggered when inserting.
    
    \item[--] \emph{Many Relabel} case: we randomly choose 1,000 positions from 10 million items in $\od$, so that each position averagely has {10,000} inserted items. 
    Thus, is it possible that {\emph{many}} \texttt{Relabel} procedures are triggered when inserting. 
    
    \item[--] \emph{Max Relabel} case: we only choose a single position (at the middle of $\od$) to insert {$10,000,000$} items. In this way, we obtain a {\emph{maximum}} number of triggered relabel procedures. 
\end{itemize}

All items are inserted on-the-fly without preprocessing. In other words, 10 million items are randomly assigned to multiple workers, e.g 32 workers, even if in the \emph{Max} case all insertions are reduced to sequential execution.

\subsection{Evaluating Relabelling}
In this test, we evaluate the \texttt{Relabel} procedures that is triggered by \texttt{Insert} operations over four test cases, \emph{No}, \emph{Few}, \emph{Many} and \emph{Max}. 
For this, different numbers of workers will have the same trend, so we choose 32 workers for this evaluation. 

\begin{table}[!htb]
\centering
%\small
\begin{tabular}{ l|r r r r|c }
    \hline
  & \multicolumn{4}{|c|}{\emph{Insert}} & \multicolumn{1}{|c}{\emph{Mixed}} \\ %\cline{2-5} %\hline 
    \footnotesize Case  & \footnotesize {Relabel}\# & $L_b$\# & $L^t$\# & \footnotesize AvgLabel\# & \footnotesize OrderRedo\# \\
     \hline
No & 0 & 10,000,000 & 0 & 1 & 0 \\
Few & 2,483 & 10,069,551 & 4,967 & 1 & 0 \\
Many & 356,624 & 19,985,472 & 5,754,501 & 2.6 & 0 \\
Max & 357,142 & 19,999,976 & 99,024,410 & \textbf{11.8} & 0 \\
 \hline
\end{tabular}
 \caption{The detailed numbers of the relabel procedure.}
 \label{tab:relabel}
\end{table}

In Table \ref{tab:relabel}, columns 2 - 4 show the details in the \emph{Insert} experiment, where \emph{Relabel\#} is the times of triggered \texttt{Relabel} procedures, \emph{$L_b$\#} is the number of updated bottom-labels for items, \emph{$L^t$\#} is the number of updated top-labels for items, and \emph{AvgLabel\#} is the average number of updated labels for each inserted items when inserting 10 million items. We can see that, for four cases, the amortized numbers of updated labels increase slowly, where the average numbers of inserted items for each position increase by $1$, $10$, $10$ million, and $10$ billion. This is because our parallel \texttt{Insert} operations have $O(1)$ amortized work.

\iftrue
\begin{itemize} %[leftmargin=*,noitemsep]
    \item[--] The \emph{No} case not triggers \texttt{Relabel}, updating only one $L_b$ per insert. 
    
    \item[--] The \emph{Few} case triggers $2.5$ thousand \texttt{Relabel}, updating $1.007$ $L_b$, $0.005$ $L^t$, and totally about one labels per inserted items.
    
    \item[--] The \emph{Many} case triggers $0.36$ million \texttt{Relabel}, updating $2$ $L_b$, $0.6$ $L^t$, and totally about $2.6$ labels per inserted items. 

    \item[--] The \emph{Max} case triggers $0.36$ million \texttt{Relabel}, which is the same as \emph{Many} the case. But, it updates $2$ $L_b$, $9.9$ $L^t$, totally about $11$ labels per inserted items.   
\end{itemize}
\fi

In Table \ref{tab:relabel}, the last column shows the numbers of \emph{redo} for \texttt{Order} operations in the \emph{Mixed} experiment, which are all zero. Since the \emph{Mixed} has mixed \texttt{Order} and \texttt{Insert} operations, we may redo the \texttt{Order} operation if the corresponding labels are being updated. However, \texttt{Relabel} happens with a low probability; also, it is a low probability that related labels are changed when comparing the order of two items. This is why the numbers of \emph{redo} are zero, leading to high parallel performance.

%%%%%%%%%%%%%%%%%%%%%%%%%%%%%%%%
% add the base-line (the sequential version)
% the compared methods. 
%%%%%%%%%%%%%%%%%%%%%%%%%%%%
\subsection{Evaluating the Running Time}
In this test, we exponentially increase the number of workers from 1 to 64 and evaluate the real running time. 
We perform \emph{Insert}, \emph{Order}, \emph{Delete}, and \emph{Mixed} over four test cases, \emph{No}, \emph{Few}, \emph{Many} and \emph{Max}. 

 \begin{figure*}[!htb]
      \centering
      % include first image
      \includegraphics[width=0.7\linewidth]{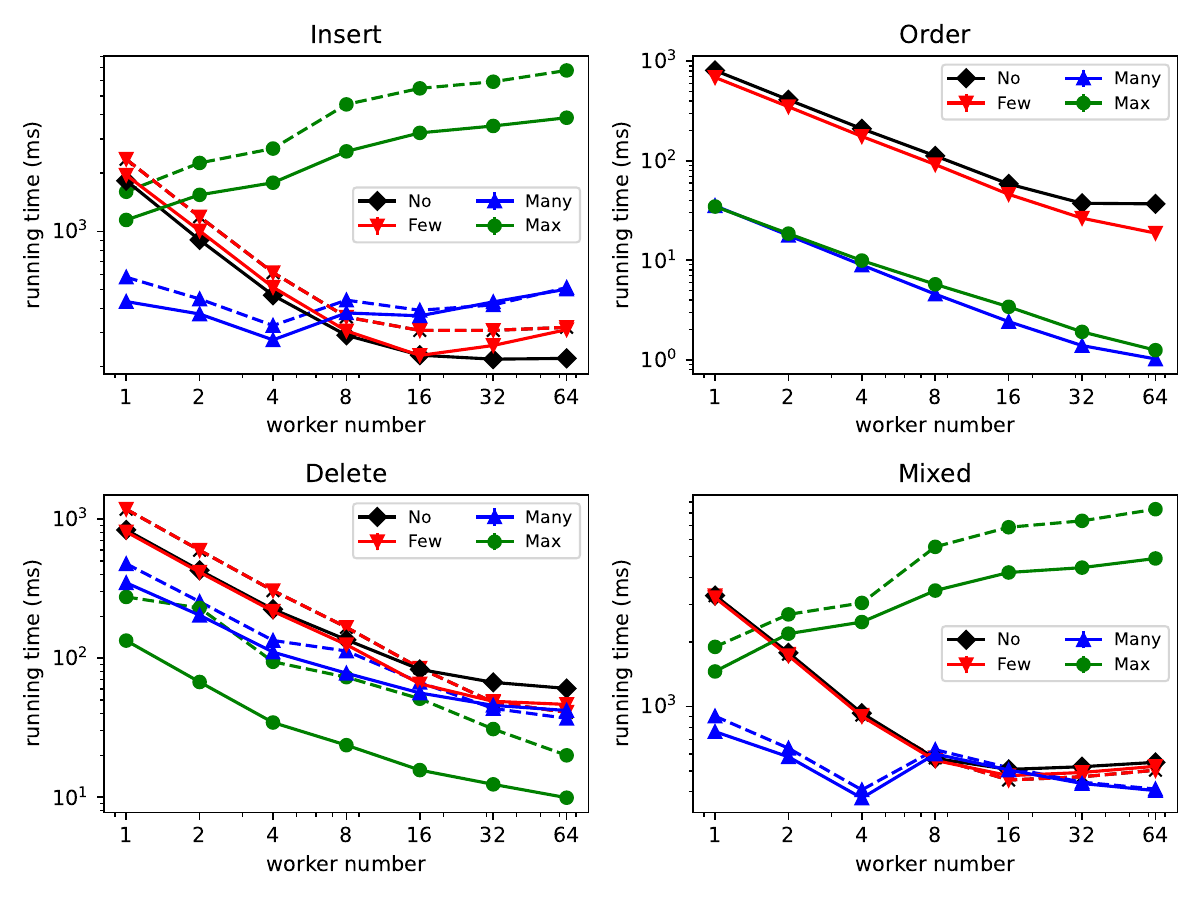}  
      \caption{Evaluating the running times.}
      \label{fig:sub-first}
\end{figure*}

\iffalse 
\begin{figure*}[!htb]
    
    \begin{subfigure}{\textwidth}
      \centering
      % include first image
      \includegraphics[width=0.7\linewidth]{om-time-2.pdf}  
      \caption{The running times}
      \label{fig:sub-first}
    \end{subfigure}
    
    \begin{subfigure}{\textwidth}
      \centering
      % include second image
      \includegraphics[width=0.7\linewidth]{om-speedup-2.pdf}  
      \caption{The speedups}
      \label{fig:sub-second}
    \end{subfigure}

    \begin{subfigure}{\textwidth}
      \centering
      % include third image
    \includegraphics[width=0.7\linewidth]{om-scalability-2.pdf}  
      \caption{The scalability with $32$ workers}
      \label{fig:sub-third}
    \end{subfigure}
\caption{Evaluating the running times, speedups, and scalability.}
%\caption{Evaluating the running times and speedups.}
\label{fig:exp}
\end{figure*}
\fi

The plots in Figure \ref{fig:sub-first} depict the performance. The x-axis is the number of workers and the y-axis is the execution time (millisecond). 
Note that, we compare the performance by using two kinds of lock: the OpenMP lock (denoted by solid lines) and the spin lock (denoted by dash lines). A first look reveals that the running times normally decrease with increasing numbers of workers, except for the \emph{Max} case over \emph{Insert} and \emph{Mixed} experiments. 
Specifically, we make several observations:
\begin{itemize}%[noitemsep,topsep=0pt,leftmargin=*]%%[noitemsep,topsep=0pt,leftmargin=*]
    \item[--] Three experiments, \emph{Insert}, \emph{Delete} and \emph{Mixed}, that use the spin lock are much faster than using the OpenMP lock. This is because the lock regions always have few operations, and busy waiting (spin lock) is much faster than the suspension waiting (OpenMP lock). 
    Unlike the above three experiments, the \emph{Order} experiment does not show any differences since \texttt{Order} operations are lock-free without using locks for synchronization. 
    
    \item[--] For the \emph{Max} case of \emph{Insert} and \emph{Mixed}, abnormally, the running time is increasing with an increasing number of workers. The reason is that the \texttt{Insert} operations are reduced to sequential in the \emph{Max Case} since all items are inserted into the same position. Thus it has the highest contention on shared positions where multiple workers accessing at the same time, especially for $64$ workers. 
    
    \item[--] For the \emph{Many} case of \emph{Insert} and \emph{Mixed}, the running times are decreasing until using $4$ workers. From $8$ workers, however, the running times begin to increase. 
    This is because the \texttt{Insert} operations have only 1,000 positions in the \emph{Many} case, and thus it may have high contention on shared positions when using more than $4$ workers. 

    \item[--] Over the \emph{Order} and \emph{Delete} experiments, we can see the \emph{Many} and \emph{Max} cases are always faster than the \emph{Few} and \emph cases. This is because the \emph{Few} and \emph{No} cases have $1,000$ and $1$ operating positions, respectively; all of these positions can fit into the CPU cache with high probability, and accessing the cache is much faster than accessing the memory.
\end{itemize}

\subsection{Evaluating the Speedups}

   \begin{figure*}[!htb]
      \centering
      % include second image
      \includegraphics[width=0.7\linewidth]{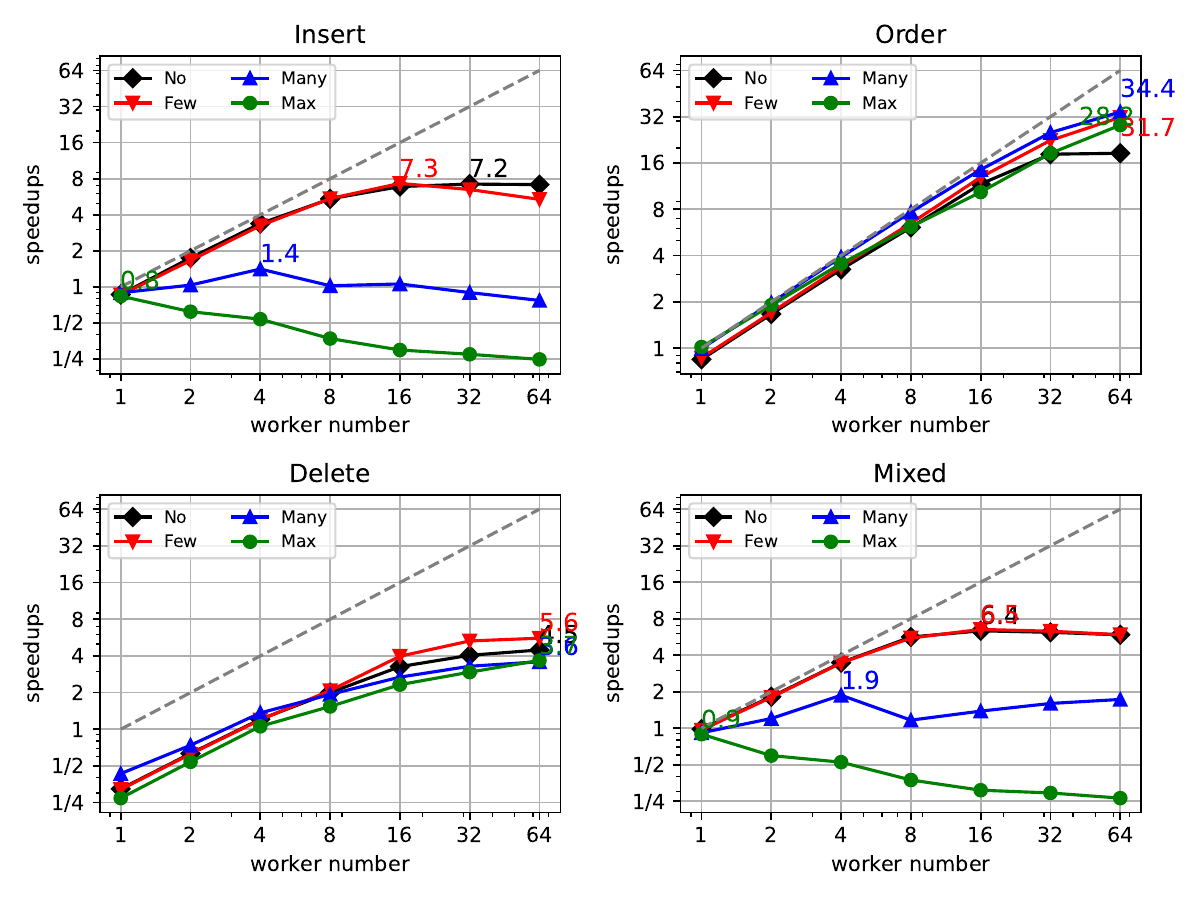}  
      \caption{Evaluate the speedups.}
      \label{fig:sub-second}
    \end{figure*}
    
The plots in Figure \ref{fig:sub-second} depict the speedups. The x-axis is the number of workers, and the y-axis is the speedups, which are the ratio of running times (by using spin locks) between the sequential version and using multiple workers. The dotted lines show the perfect speedups as a baseline. The numbers beside the lines indicate the maximal speedups. 
A first look reveals that all experiments achieve speedups when using multiple cores, except for the \emph{Max} case over \emph{insert} and \emph{Mixed} experiments.
Specifically, we make several observations:
\begin{itemize}%[noitemsep,topsep=0pt,leftmargin=*] %[noitemsep,topsep=0pt,leftmargin=*]
    % \item[--] Over \emph{Insert} and \emph{Mixed}, the \emph{Max} case achieve The reason is that the \texttt{Insert} operations has a worst case for the \emph{Max} case. That is, all items are inserted into the same position, which reduces to sequential. 
    \item[--] For all experiments, we observe that the speedups are around $1/4$ to $1$ when using $1$ worker in all cases. This is because, for all operations of OM, the sequential version has the same work as the parallel version. Especially, for \emph{Delete}, such speedups are low as $1/2$ - $1/4$, as locking items for deleting costs much running time. 
    
    \item[--] For \emph{Insert} and \emph{Mixed}, we achieve around $7$x speedups using $32$ workers in \emph{No} and \emph{Few} cases, and around $2$x speedups using $4$ workers in \emph{Many} cases.
    This is because all CPU cores have to access the shared memory by the bus, which connects memory and cores, and the atomic \texttt{CAS} operations will lock the bus. 
    Each \texttt{Insert} operation may have many atomic \texttt{CAS} operations for spin lock and many atomic read and write operations for updating labels and lists.  
    In this case, the bus traffic is high, which is the performance bottleneck for \emph{Insert} operations.
    
    \item[--] For \emph{Order}, all four cases achieve almost perfect speedups from using $1$ to $32$ workers, as \texttt{Order} operations are lock-free. 

    \item[--] For \emph{Delete}, it achieves around $4$x speedups using $64$ workers in four cases.
    This is because, for parallel \texttt{Delete} operations, the worst case, which is all operations are blocking as a chain, is almost impossible to happen.  
\end{itemize}

% \begin{itemize}[noitemsep,topsep=0pt,leftmargin=*]
%     \item Best-case (BC for short): we perform all operations randomly over 1 million initial items in $\od$, by which all workers may have low probability on lock contention.
%     \item Average-case (AC for short): we perform all operations randomly over $1000$ consecutive items in the middle of $\od$, by which all workers may have much higher probability on lock contention than the best-case.
%     \item Worst-case (WC for short): we perform all operations over a fixed item in $\od$, by which all workers may have lock contention and reduce to sequential.  
% \end{itemize}

%Show the table of speedups with 64 workers.

\iftrue
\subsection{Evaluating the Stability}
In this test, we compare 100 testing times for the \emph{Insert}, \emph{Order}, and \emph{Delete} operations by using 32 workers. 
Each time, we randomly choose positions and randomly insert items for the \emph{NO}, \emph{Few} and \emph{Many} cases, so that the test is different. However, it is always the same for the \emph{Max} case, since there is only one position to insert all items.  

\iftrue
\begin{figure*}[!htb]
\centering
      \includegraphics[width=0.7\linewidth]{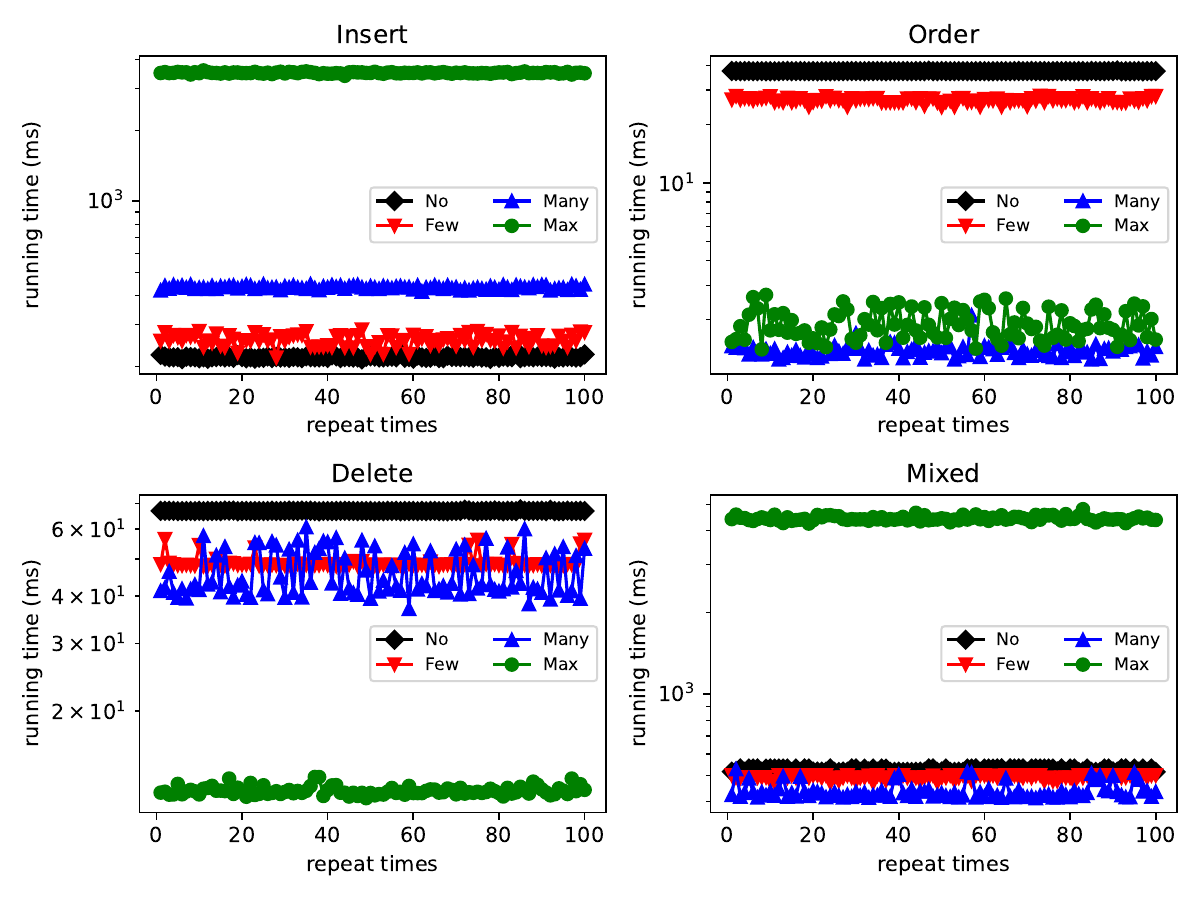}  
    \caption{Evaluating the stability of running times over 32 cores.}
    \label{fig:stability}
\end{figure*}
\fi

The plots in Figure \ref{fig:stability} depict the running time by performing the experiments 100 times. 
The x-axis is the index of repeating times and the y-axis is the running times (millisecond). 
We observe that the performance of \emph{Insert}, \emph{Order}, \emph{Delete} and \emph{Mixed} are all well bounded for all four cases. 
We observe that the \emph{Max} case has wider variation then other cases over \emph{Insert} and \emph{Mixed}. This is because the parallel \texttt{Insert} operations always have contention on shared data in memory. Such contention causes the running times to variate within a certain range. 
\fi

\iftrue
%%% all the other experiment
\subsection{Evaluating the Scalability}
In this test, we increase the scale of the initial order list from 10 million to 100 million and evaluate running times with fixed $32$ workers. We test three cases, \emph{No}, \emph{Few}, and \emph{Many}, by fixing the average number of items per insert position. For example, given an initial order list with 20 million items, the \emph{No} case has 20 million insert positions, the \emph{Few} case has 2 million positions, and the \emph{Many} case has 2,000 insert positions. Since the \emph{Max} case is reduced to sequential and can be optimized by using a single worker, we omit it in this test. 

 \begin{figure*}[!htb]
      \centering
      % include third image
    \includegraphics[width=0.7\linewidth]{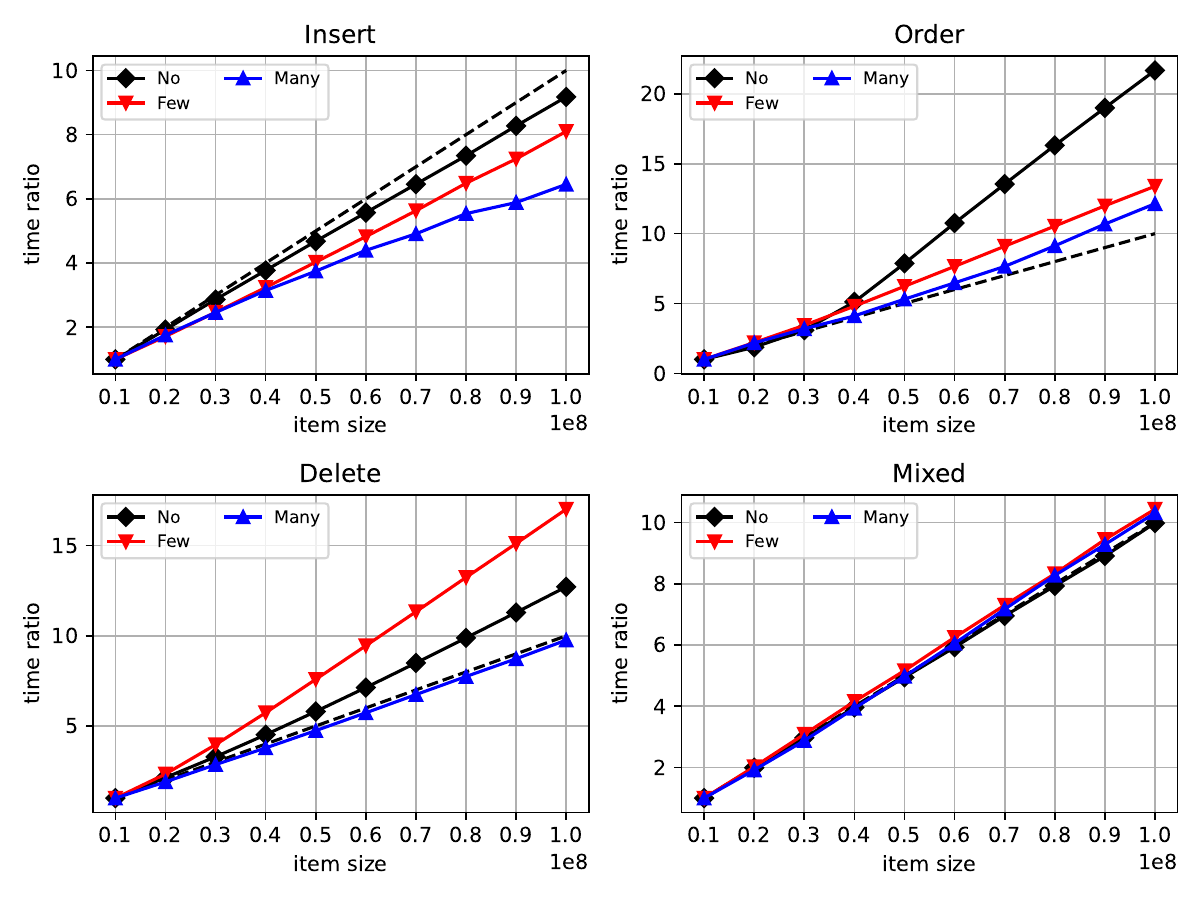}  
      \caption{Evaluate the scalability with $32$ workers}
      \label{fig:sub-third}
    \end{figure*}

The plots in Figure \ref{fig:sub-third} depict the performance. The x-axis is the initial size of the order list, and the y-axis is the time ratio of the current running time to the ``10 million'' running time. 
The dotted lines show the perfect time ratio as a baseline.
Obviously, the beginning time ratio is one. We observe that the time ratios are roughly close to linearly increasing with the scales of the order list. This is because all parallel \texttt{Insert}, \texttt{Delete}, and \texttt{Order} have best-case time complexity $O(\frac{m}{\mathcal P})$ and averagely their running times are close to the best case.

Specifically, for \emph{Order}, we can see the time ratio is up to 20x with a scale of 100 million in \emph{No} case. This is because \emph{No} case has 100 million positions for random \texttt{Order} operations, which is not cache friendly; also, by increasing the scale of data, the cache hit rate decreases, so the performance is affected.

% Given a ordered list $\od$ with 1,000 items, we we exponentially increase the number of operations from 1,000 to 100,000,000 with one worker, and 64 workers, respectively. The operations are mixed averagely with \texttt{Order}, \texttt{Insert}, \texttt{Delete}. If there are not enough items can be deleted, the delete operation can be delayed until new items are inserted.

% Also cover three cases. 
\fi

\section{Conclusions and Future Work}
We present a new parallel order maintenance (OM) data structure. The parallel \texttt{Insert} and \texttt{Delete} are synchronized with locks efficiently. Notably, the parallel \texttt{Order} is lock-free, and can execute highly in parallel. Experiments demonstrate significant speedups (for 64 workers) over the sequential version on a variety of test cases. 

In future work, we will attempt to reduce the synchronization overhead, especially for parallel \texttt{Insert}. We will investigate insertions and deletions in batches by preprocessing the inserted or deleted items, respectively, which can reduce the contention for multiple workers. 
We also intend to apply our parallel OM data structure to many applications.

\iffalse 
%% Acknowledgments
\begin{acks}                            %% acks environment is optional
                                        %% contents suppressed with 'anonymous'
  %% Commands \grantsponsor{<sponsorID>}{<name>}{<url>} and
  %% \grantnum[<url>]{<sponsorID>}{<number>} should be used to
  %% acknowledge financial support and will be used by metadata
  %% extraction tools.
  This material is based upon work supported by the
  \grantsponsor{GS100000001}{National Science
    Foundation}{http://dx.doi.org/10.13039/100000001} under Grant
  No.~\grantnum{GS100000001}{nnnnnnn} and Grant
  No.~\grantnum{GS100000001}{mmmmmmm}.  Any opinions, findings, and
  conclusions or recommendations expressed in this material are those
  of the author and do not necessarily reflect the views of the
  National Science Foundation.
\end{acks}
\fi

%\bibliographystyle{unsrt}
\bibliographystyle{elsarticle-num}
\bibliography{references}
\end{document}